\documentclass[11pt]{article}

\usepackage{amsmath}
\usepackage{amsthm}
\usepackage{amsfonts}
\usepackage{amssymb}
\usepackage{latexsym} 
\usepackage{epsfig}
\usepackage{graphicx}
\usepackage{calc}  %%required to make lists work correctly
\usepackage[matrix,tips,graph,curve,web]{xy}

\makeatletter

\def\@maketitle{%
  \newpage
  \null
  \let \footnote \thanks
    {\normalfont\sffamily\bfseries\Large\noindent\@title \par}%
    \vskip 1em%
    {\normalfont\sffamily %\large
        \noindent
        \@author
        \par}
  \par
  \vskip 4em}
\def\@seccntformat#1{\csname the#1\endcsname{.\ }}
\renewcommand\section{\@startsection {section}{1}{\z@}%
                                   {-3.0ex \@plus -1ex \@minus -.2ex}%
                                   {1.5ex \@plus.2ex}%
                                   {\normalfont\large\bfseries}}
\renewcommand\subsection{\@startsection{subsection}{2}{\z@}%
                                     {-2.75ex\@plus -1ex \@minus -.2ex}%
                                     {1.5ex \@plus .2ex}%
                                   {\normalfont\large}}
\def\fnum@figure{\normalfont\footnotesize\figurename~\thefigure}

\renewcommand\tableofcontents{%
    \section*{\contentsname
        \@mkboth{%
           \MakeUppercase\contentsname}{\MakeUppercase\contentsname}}%
    \@starttoc{toc}%
    }
\renewcommand*\l@part[2]{%
  \ifnum \c@tocdepth >-2\relax
    \addpenalty\@secpenalty
    \addvspace{2.25em \@plus\p@}%
    \begingroup
      \setlength\@tempdima{3em}%
      \parindent \z@ \rightskip \@pnumwidth
      \parfillskip -\@pnumwidth
      {\leavevmode
       \large \bfseries #1\hfil \hb@xt@\@pnumwidth{\hss #2}}\par
       \nobreak
       \if@compatibility
         \global\@nobreaktrue
         \everypar{\global\@nobreakfalse\everypar{}}%
      \fi
    \endgroup
  \fi}
\renewcommand*\l@section[2]{%
  \ifnum \c@tocdepth >\z@
    \addpenalty\@secpenalty
    \addvspace{1.0em \@plus\p@}%
    \setlength\@tempdima{1.5em}%
    \begingroup
      \parindent \z@ \rightskip \@pnumwidth
      \parfillskip -\@pnumwidth
      \leavevmode \sffamily\bfseries
      \advance\leftskip\@tempdima
      \hskip -\leftskip
      #1\nobreak\hfil \nobreak\hb@xt@\@pnumwidth{\hss #2}\par
    \endgroup
  \fi}
\renewcommand*\l@subsection{\sffamily\@dottedtocline{2}{1.5em}{2.3em}}
\renewcommand*\l@subsubsection{\@dottedtocline{3}{3.8em}{3.2em}}
\renewcommand*\l@paragraph{\@dottedtocline{4}{7.0em}{4.1em}}
\renewcommand*\l@subparagraph{\@dottedtocline{5}{10em}{5em}}
\makeatother


\theoremstyle{plain}
\newtheorem{theorem}{Theorem}[section]
\newtheorem{corollary}[theorem]{Corollary}
\newtheorem{lemma}[theorem]{Lemma}

\newtheorem{conjecture}[theorem]{Conjecture}

\theoremstyle{definition}
\newtheorem{definition}[theorem]{Definition}

  {\begin{list}{}%
    {%
    \settowidth{\labelwidth}{#1}%
    \setlength{\itemindent}{0pt}%
    \setlength{\labelsep}{1em}%
    \setlength{\leftmargin}{\labelwidth+\parindent+\labelsep}%
    \setlength{\itemsep}{0pt}%
    \setlength{\parsep}{.6ex}}}%
  {\end{list}}

  {\begin{list}{}%
    {%
    \settowidth{\labelwidth}{#1}%
    \setlength{\itemindent}{0pt}%
    \setlength{\labelsep}{1em}%
    \setlength{\leftmargin}{\labelwidth+\labelsep}%
    \setlength{\itemsep}{0pt}%
    \setlength{\parsep}{.6ex}}}%
  {\end{list}}

\makeatletter
\newenvironment{subequations*}{% same thing but without incrementing
  \begingroup % conservative approach
  \let\protect\@nx
  \edef\@tempa{\def\@nx\theparentequation{\theequation}}%
  \@xp\endgroup\@tempa
  \setcounter{parentequation}{\value{equation}}%
  \setcounter{equation}{0}%
  \def\theequation{\theparentequation\alph{equation}}%
  \ignorespaces
}{%
  \setcounter{equation}{\value{parentequation}}%
  \global\@ignoretrue
}

\newcommand{\sgn}{\rm sgn\,}
\renewcommand\det{{\rm det\,}}

\def\d/{/\mspace{-6.0mu}/}

\usepackage{helvet}
\usepackage{graphicx,amssymb,amsmath,amsfonts,amsthm,color,url, placeins}
\usepackage{wrapfig}

\renewcommand\section{\@startsection{section}{1}{\z@}%
                                   {-3.0ex \@plus -1ex \@minus -.2ex}%
                                   {1.5ex \@plus.2ex}%
                                   {\normalfont\sffamily\large\bfseries}}
\renewcommand\subsection{\@startsection{subsection}{2}{\z@}%
                                     {-2.75ex\@plus -1ex \@minus -.2ex}%
                                     {1.5ex \@plus .2ex}%
                                   {\normalfont\sffamily\large}}
\renewcommand\subsubsection{\@startsection{subsubsection}{3}{\z@}%
                                     {-2.75ex\@plus -1ex \@minus -.2ex}%
                                     {1.5ex \@plus .2ex}%
                                   {\normalfont\sffamily\large}}

\newcommand{\od}{\stackrel{\mbox {\tiny {def}}}{=}}

\def\RR{\mathbb{R}}

\def\d{\mathrm{d}}

\def\RR{\mathbb{R}}

\def\RR{\mathbb{R}}

\def\M{\mathcal{M}}

\def\det{{\operatorname{det}}}

\def\max{\mathrm{max}}

\def\od{\stackrel{\mathrm{def}}{=}}

\def\supp{\operatorname{supp}}

\def\FP{\operatorname{FP}}
\def\idx{\operatorname{idx}}
\def \sgn{\operatorname{sgn}}
\def\B{\mathcal{B}}
\def\M{\mathcal{M}}

\usepackage{color}
\definecolor{cherry}{rgb}{0.9,.1,.2}
\definecolor{green}{rgb}{.2,.7,0}
\definecolor{blue}{rgb}{0,.0, .81}

\def\rkatie#1{{#1}}

\begin{document}

\noindent {\Large \bf Diversity of emergent dynamics in competitive threshold-linear networks}\\

\noindent Katherine Morrison$^{1}$, Anda Degeratu$^{2}$, Vladimir Itskov$^{3}$, Carina Curto$^{3}$\\ 
October 14, 2023\\

\begin{small}
\noindent $^1$ School of Mathematical Sciences, University of Northern Colorado, Greeley, CO 80639\\
$^2$ Fachbereich Mathematik, Universität Stuttgart, Stuttgart, Germany, 70569\\
$^3$ Department of Mathematics, The Pennsylvania State University, University Park, PA 16802\\
\end{small}

\vspace{-.25in}

\paragraph{Abstract.}
Threshold-linear networks consist of simple units interacting in the presence of a threshold nonlinearity.  Competitive threshold-linear networks have long been known to exhibit multistability, where the activity of the network settles into one of potentially many steady states.  In this work, we find conditions that guarantee the {\it absence} of steady states, while maintaining bounded activity.   These conditions lead us to define a combinatorial family of competitive threshold-linear networks, parametrized by a simple directed graph.  By exploring this family, we discover that threshold-linear networks are capable of displaying a surprisingly rich variety of nonlinear dynamics, including limit cycles, quasiperiodic attractors, and chaos.  In particular, several types of nonlinear behaviors can co-exist in the same network.  Our mathematical results also enable us to engineer networks with multiple dynamic patterns.  Taken together, these theoretical and computational findings suggest that threshold-linear networks may be a valuable tool for understanding the relationship between network connectivity and emergent dynamics.

\tableofcontents

\section{Introduction}
Networks are complex dynamical systems that consist of nodes and their interactions.  They are commonly-used models in fields as disparate as ecology, economics, and neuroscience.  Even when the building blocks are simple, networks can display rich {\it emergent dynamics}, whose complexity cannot be reduced to a sum of constituent parts.  Moreover, the most interesting dynamic phenomena that arise are fundamentally nonlinear behaviors, such as multistability, periodic attractors, and chaos.  

Despite this, network dynamics are often approximated using linear models -- namely, linear systems of ordinary differential equations (ODEs).  This is because the accompanying mathematical theory is well developed.  Indeed, one might say that networks with linear interactions are the complex systems we already understand.
While reducing more complicated models to linear approximations can be a useful approach, this strategy also poses severe limitations.  Phenomena such as multistability, chaos, and robust periodic attractors (limit cycles) simply do not occur in linear models.  Can we replace approximation by linear systems of ODEs
with something ``almost'' linear -- simple enough that a useful mathematical theory can be developed, yet capable of capturing the full variety of nonlinear behavior? 

Motivated by this question, we study the dynamics of threshold-linear networks (TLNs).  Emergent dynamics in these networks are not inherited from intrinsically oscillating nodes or a fluctuating external input -- instead, they can be attributed solely to the structure of connectivity, given by a matrix $W$. 
The nonlinear behavior stems entirely from the presence of a simple threshold at each node, which guarantees that the activity of individual units cannot go negative. This nonnegativity is natural in any system where the dynamic variables represent fundamentally nonnegative quantities, such as the size of a population, a chemical concentration, or the firing rate of a neuron.  Though TLNs look essentially linear, the presence of the threshold changes everything.  With it, the entire repertoire of complex nonlinear behavior comes into play: multistability, limit cycles, quasi-periodic behavior, and even deterministic chaos emerges.  

\rkatie{Historically, TLNs have been studied with an eye towards {\em stable fixed points}, as these are the traditional attractors considered in the Hopfield model and related neural network literature \cite{Hopfield1,Hopfield2}. This led to an early emphasis on the case of symmetric $W$, where convergence to stable fixed points can be guaranteed \cite{HahnSeungSlotine,XieHahnSeung}. In the early 2000s, the study of ``permitted'' and ``forbidden'' sets also provided simple conditions under which symmetric (inhibitory) TLNs give rise to multistability, and showed that the collection of all permitted sets of a symmetric network satisfies the structure of a simplicial complex \cite{HahnSeungSlotine}. Notably, a permitted set is not a fixed point but a subset of neurons that can in principle be co-activated at a stable fixed point for at least one external input; this simplifies the problem to one of analyzing the spectral properties of principal submatrices of $W$, while ignoring questions about which fixed points can arise for a given external input.}

\rkatie{The present authors continued the study of stable fixed points and permitted sets of TLNs in subsequent work \cite{flex-memory,net-encoding,pattern-completion}, also focusing primarily on the symmetric case. We then shifted attention to non-symmetric $W$, leading to the work presented here. A preliminary (unpublished) version of the current article was our first serious foray into the dynamics of non-symmetric TLNs \cite{CTLN-prelim}; those results became the seeds of this article and inspired related developments \cite{fp-paper,book-chapter,sequential-att-paper,Horacio-paper,core-motifs}. In addition to biological realism, our primary motivation for this shift was a desire to explore a wider range of nonlinear behavior, including more complex attractors. There is also, however, a body of related work in the non-symmetric case that maintains its focus on stable fixed points \cite{nozari2018stability,nozari2020hierarchical}.}

\rkatie{In order to identify networks with complex attractors, we initially sought 
conditions that would guarantee the {\it absence} of stable fixed points, so that network activity would be forced to be oscillatory or chaotic.  This resulted in Theorem~\ref{thm:Thm1}, whose conditions on pairwise interactions can be framed in terms of a directed graph. These conditions then led us to define a combinatorial family of competitive threshold-linear networks that are parametrized by a directed graph; we call this family the {\it CTLN model} (``C" for ``combinatorial").  By exploring CTLNs, we have discovered that TLNs are capable of displaying a surprisingly rich variety of nonlinear dynamics, which we illustrate in Section~\ref{sec:simulations}.}

Although they exhibit high-dimensional nonlinear dynamics, TLNs and CTLNs are also surprisingly tractable. We have been able to prove a series of mathematical results: Theorems~\ref{thm:parity},~\ref{thm:Thm1},~\ref{thm:Thm1b}, and~\ref{thm:Thm2}, which provide valuable information about the stable and unstable fixed points of these networks. \rkatie{Theorem~\ref{thm:parity} is an index theorem, which constrains the set of fixed points that can coexist in a given TLN with fixed input. Specifically, for a TLN with connectivity matrix $W$, each fixed point $x^*$ has an {\em index}, $\idx(\sigma) = \sgn \det(I-W_\sigma) \in \{\pm 1\}$, where $\sigma \subseteq [n]$ is the set of active neurons of $x^*$. Theorem~\ref{thm:parity} states that these indices must sum to $+1$. In particular, this implies the total number of fixed points is odd. Since the index of a stable fixed point is always $+1$, we also obtain an upper bound on the total number of stable fixed points.}

\rkatie{As previously discussed, Theorem~\ref{thm:Thm1} provides conditions guaranteeing the {\em absence} of stable fixed points in TLNs with bounded activity. Because the conditions are graph-theoretic, and independent of the precise choice of weights $W$, it is easy to specialize this result to the case of CTLNs, yielding Theorem~\ref{thm:Thm1b}.  
Finally, Theorem~\ref{thm:Thm2} revisits the topic of stable fixed points, identifying graph structures that give rise to stable equilibria in CTLNs.}

\rkatie{The proofs of Theorems~\ref{thm:parity},~\ref{thm:Thm1},~\ref{thm:Thm1b}, and~\ref{thm:Thm2} required novel approaches and different techniques from what was previously used in the symmetric case. In particular, in Section~\ref{sec:proofs} we consider TLNs as a patchwork of linear systems, and develop methods to carefully analyze the relationship between fixed points of the component linear systems and those of the full nonlinear TLN. The proof of Theorem~\ref{thm:parity} involves a novel application of the Poincaré-Hopf theorem; to our knowledge, this has not previously been used in the TLN literature. We also prove a number of new lemmas, specifically for the non-symmetric case. For example, a key result in proving Theorem~\ref{thm:Thm1} is  Lemma~\ref{lemma:2x2}, which relates the stability of a non-symmetric matrix to that of its $2 \times 2$ principal submatrices.}

\rkatie{What is perhaps most striking about our CTLN theorems is how they have enabled us to engineer networks exhibiting multiple dynamic attractors, with distinct attractors corresponding to different initial conditions. In Section~\ref{sec:simulations} we illustrate the diversity of emergent dynamics via simulations of example networks that are guaranteed to have no stable fixed points (by Theorem~\ref{thm:Thm1b}). We also use Theorem~\ref{thm:Thm2} to generate networks with transiently active ``cell assemblies."
We conclude this section with some ``engineered'' networks exhibiting multiple prescribed dynamic attractors. In each of these cases, the networks are built using constraints and intuition relating directly to the graph of connectivity, without any need for parameter tuning.}  Taken together, our theoretical and computational results suggest that TLNs are a valuable tool for studying the relationship between emergent dynamics and network connectivity.

\rkatie{While CTLNs are a subfamily of TLNs, it should be noted that TLNs are themselves a special case of a much larger class of (non-smooth) switching dynamical systems. These systems are piecewise-linear, and have been extensively studied in the context of stability and control (see e.g. \cite{Liberzon2003,Johansson2003,LinAntsaklis2005,pavlov2005convergent,Bernardo2008}). 
Interestingly, by studying the constrained subfamily of CTLNs we have been able to obtain a richer variety of dynamics, as well as more detailed theoretical results, than what is readily available in the more general piecewise-linear settings. Nevertheless, it would be valuable to better understand how the results presented here relate to the broader switching dynamical systems literature.} 

\section{Preliminaries}\label{sec:prelim}

A {\it threshold-linear network} (TLN) is a rate model consisting of $n$ nodes, with dynamics governed by the system of ordinary differential equations:
\begin{equation}\label{eq:network}
\dfrac{dx_i}{dt} = -x_i + \left[\sum_{j=1}^n W_{ij}x_j+b_i \right]_+, \quad i = 1,\ldots,n.
\end{equation}
The dynamic variables $x_1,\ldots,x_n$ give the activity levels\footnote{If the nodes are neurons, the activity level is typically called a `firing rate.'} of nodes $1,\ldots,n$.
The matrix entries $W_{ij}$ are directed connection strengths between pairs of nodes, the vector $b=(b_1,\ldots,b_n) \in \RR^n$ represents the external drive to each node, and the threshold-nonlinearity $[\cdot]_+$ is given by $[y]_+ \od \max\{y,0\}$. We refer to the TLN with matrix $W$ and vector $b$ as $(W,b)$.  \rkatie{Note that although TLNs are nonlinear dynamical systems, they are piecewise linear: the threshold-nonlinearity decomposes the state space into chambers, within which the dynamics are linear, of the form $dx/dt= Ax+c$ (see Section~\ref{sec:linear-systems} for more details).}

A {\it fixed point} of~\eqref{eq:network} is a vector $x^* \in \RR^n$ that satisfies $\left.\dfrac{dx_i}{dt}\right|_{x=x^*} = 0$ for each $i \in \{1, \ldots, n\}$. We will be interested in both stable and unstable fixed points of TLNs.  \rkatie{The \emph{support} of a fixed point $x^*$ is the subset of active nodes, 
\begin{equation}\label{eq:supp}
\supp{x^*} \od \{i \mid x^*_i>0\}.
\end{equation}
}

\rkatie{ Throughout this paper, we restrict ourselves to considering TLNs that are \emph{nondegenerate} (a precise definition is given in Definition~\ref{def:nondegenerate}).   Briefly, nondegeneracy requires that certain determinants are non-zero, such that each of the linear systems of the TLN is nondegenerate and thus has a unique fixed point.   Another requirement is that at least one $b_i>0$, which guarantees that $0$ is never a fixed point of the TLN. A key feature of nondegenerate TLNs is that they are guaranteed to have at most one fixed point per support (see Corollary~\ref{cor:fp-bound}).  Note that almost all networks of the form~\eqref{eq:network} are nondegenerate, since having a zero determinant is a highly fine-tuned condition. }

Although these networks have been around for decades in the neural networks community, the mathematical theory is still a work in progress.  It began in earnest about 20 years ago, with work by Hahnloser, Seung, and others \cite{HahnSeungSlotine, XieHahnSeung, Hahn2000}.  Theirs was the first serious attempt to develop a mathematical theory of threshold-linear networks to rival that of Hopfield networks \cite{Hopfield1}.  Not surprisingly, the initial results were confined to the case where $W$ is a symmetric matrix.  In \cite{HahnSeungSlotine}, precise conditions were found to guarantee that network activity always converges to a stable fixed point, and a characterization was given of symmetric threshold-linear networks exhibiting multistability. In \cite{XieHahnSeung}, the role of lateral inhibition was highlighted as playing an especially important role, enabling selective competition between groups of neurons.

\subsection{Competitive TLNs}   
A {\it competitive} threshold-linear network is a special case of~\eqref{eq:network} with the additional restrictions: $W_{ij} \leq 0$, $W_{ii} = 0$, and $b_i \geq 0$ for all $i,j = 1,\ldots,n$.  \rkatie{\emph{Unless otherwise specified, we will assume all TLNs are both {\bf competitive} and {\bf nondegenerate}.} Note that nondegeneracy implies that not all $b_i$ can be 0, and so despite the competitive dynamics, trajectories cannot decay to the origin.}

An important property of competitive networks is that they are guaranteed to have bounded activity.  In fact, we can prove that the asymptotic dynamics are confined to the box,
$$\B \od \prod_{i=1}^n [0,b_i].$$

\begin{lemma}\label{lemma:bounded}
 The box $\B = \prod_{i=1}^n [0,b_i]$ is a globally attracting set of the \rkatie{competitive} TLN dynamics~\eqref{eq:network}. That is, if $x(0) \in \B$, then $x(t) \in \B$ for all $t>0$. Moreover, if $x(0) \not\in \B,$ the trajectory $x(t)$ approaches or enters $\B$ as $t \to \infty$.
\end{lemma}

\begin{proof}
To show that trajectories that start inside $\B$ stay there, it suffices to show that for each $x_i,$ the derivative $dx_i/dt$ on the boundary $\partial \B$ is either $0$ or points inside the box. That is, for $x \in \B$ we must have $dx_i/dt \geq 0$ whenever $x_i = 0$, and $dx_i/dt \leq 0$ whenever $x_i = b_i$. Now observe that $x \in \B$ implies $x_i \geq 0$ for each $i$; putting this together with the condition that $W_{ij} \leq 0$ we have
$$\left[\sum_{j=1}^n W_{ij}x_j+b_i \right]_+ \leq b_i.$$
From here we immediately see that for all $x \in \B$ (in fact, for all $x \geq 0$),
$$-x_i \leq \dfrac{dx_i}{dt} \leq -x_i+b_i.$$
Evaluating the derivatives on the boundary, we see that $dx_i/dt \geq 0$ at $x_i = 0$,
and $dx_i/dt \leq 0$ at $x_i = b_i$. 

Next, consider what happens for $x(0) \not\in \B$. If $x_i \notin [0,b_i]$ for some $i$, then either $x_i < 0$ and thus $dx_i/dt > 0$, or $x_i > b_i$ and thus $dx_i/dt < 0$. If follows that any trajectory $x(t)$ initialized outside of $\B$ will either approach or enter $\B$ as $t \to \infty$.
\end{proof}

\rkatie{In particular, Lemma~\ref{lemma:bounded} guarantees that $\mathcal{B}$ contains all the fixed points.}

\begin{corollary}\label{cor:box-fp}
All fixed points of a \rkatie{competitive} TLN $(W,b)$ lie inside $\mathcal{B} = \prod_{i=1}^n [0,b_i]$.
\end{corollary}

\rkatie{This fact is also easy to see directly:  at a fixed point, we must have 
$x_i^* =  \left[\sum_{j=1}^n W_{ij}x_j^*+b_i \right]_+ \geq 0$ for each $i$, which immediately implies that $0 \leq x_i^*\leq b_i,$ since $W_{ij} \leq 0$.}

\subsection{The set of all fixed points $\FP(W,b)$}

\rkatie{Recall that nondegenerate TLNs have at most one fixed point per support (for the proof, see Corollary~\ref{cor:fp-bound}).} We can thus label all the fixed points of a given network by their supports. We denote this as:
%\vspace{-.05in}
\begin{eqnarray*}
\FP(W,b) &\od& \{\sigma \subseteq [n] \mid   \sigma = \supp{x^*} \text{ for some } \text{fixed point } x^* \text{ of the associated TLN} \},
\end{eqnarray*}
where $$[n] \od \{1,\ldots,n\}.$$

Note that for each support $\sigma \in \FP(W,b)$, the fixed point itself is easily recovered. Outside the support, $x_i^* = 0$ for all $i \not\in \sigma$. Within the support, $x^*$ is given by:
$$x_\sigma^* = (I-W_\sigma)^{-1} b_\sigma,$$
where $x_\sigma^*$ and $b_\sigma$ are the column vectors obtained by restricting $x^*$ and $b$ to the indices in $\sigma$, and $W_\sigma$ is the induced principal submatrix obtained by restricting rows and columns of $W$ to $\sigma$. Note that $x^* \in \B$ (see Corollary~\ref{cor:box-fp}), though it is not obvious from this above formula.

In \cite{HahnSeungSlotine}, the authors studied the collection of {\em stable} fixed points of these networks through the lens of \emph{permitted sets} -- these are subsets $\sigma \subseteq [n]$ for which the network $(W,b)$ supports a stable fixed point for at least one $b \in \RR^n$.  The theory of permitted sets was further developed in \cite{flex-memory, net-encoding}. \rkatie{In particular, it was shown that a fixed point with support $\sigma$ is stable if and only if all eigenvalues of $(-I+W)_\sigma$ have negative real part (this was first shown for symmetric matrices in \cite{HahnSeungSlotine}, then generalized in \cite{net-encoding}).} Finally, in \cite{pattern-completion}, attention was shifted to the study of stable fixed points that can be simultaneously realized for a single uniform external input $b$.   In this context, there is additional mathematical tractability that allows us to generalize to asymmetric networks. Thus, we will pay special attention to the case of uniform inputs, where $b_i = \theta > 0$ for all $i$. In order to further isolate the role of network connectivity in shaping dynamics, we will also consider a special family of TLNs, known as CTLNs. These are networks where the matrix $W$ has binary synapses, corresponding to strong and weak inhibition, dictated by an underlying connectivity graph.

\subsection{The CTLN model}
The self-inhibition (or decay rate) of a single node, which has been normalized to $-1$, provides a natural scale for the strength of inhibition. If $W_{ij}> -1$, then node $j$ inhibits node $i$ {\em less} than it inhibits itself. If, on the other hand, $W_{ij} < -1$, the inhibition from $j$ to $i$ is stronger than the self-inhibition of $j$. This distinction ends up playing an important role in shaping the dynamics.

To simplify our study, and further isolate the role of connectivity, we specialize
 to competitive networks with only two values for the connection strengths: one value for $W_{ij} < -1$, and another for $W_{ij}>-1$.  To any simple directed graph $G$ on $n$ vertices, and for any $0<\varepsilon <1 $ and $\delta > 0$, we can associate a corresponding $n \times n$ connectivity matrix $W = W(G,\varepsilon,\delta)$ as follows:
\begin{equation} \label{eq:binary-synapse}
W_{ij} = \left\{\begin{array}{cc} 0 & \text{ if } i = j, \\ -1 + \varepsilon & \text{ if } i \leftarrow j \text{ in } G,\\ -1 -\delta & \text{ if } i \not\leftarrow j \text{ in } G, \end{array}\right.
\end{equation}
where $i \leftarrow j$ indicates that $G$ has an edge from $j$ to $i$, and $i \not\leftarrow j$ indicates there is no such edge. Clearly, $W$ satisfies the conditions of a competitive network.

\begin{figure}[!ht]
\begin{center}
\vspace{.1in}
\includegraphics[width=5.75in]{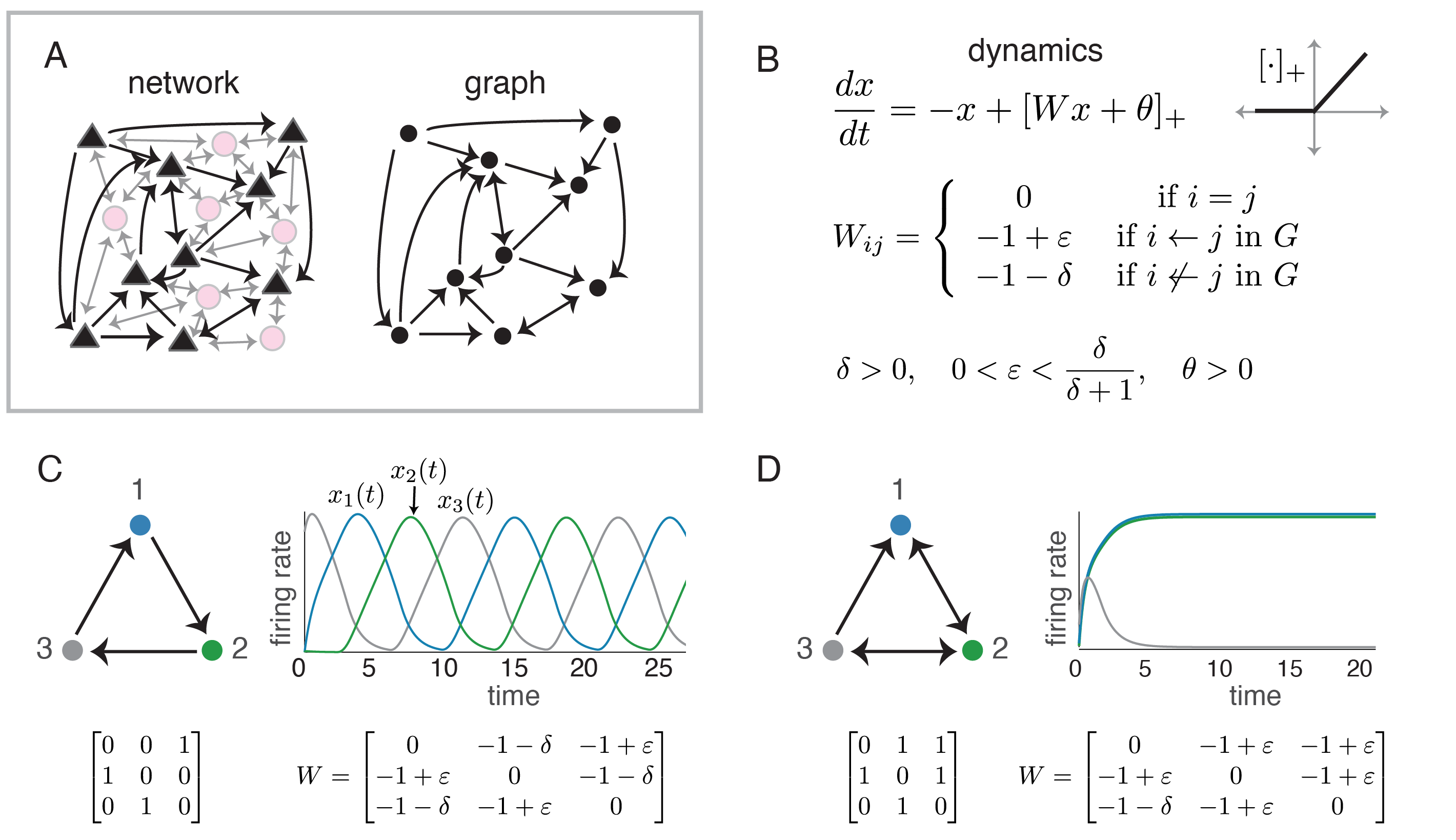}
\vspace{.1in}
\caption{{\bf Combinatorial threshold-linear networks with strong and weak inhibition.}  {(A)} (Left) A neural network with excitatory pyramidal neurons (triangles) and a background network of inhibitory interneurons (pink circles) that produce a global inhibition. (Right) The corresponding graph retains only the excitatory neurons and their connections.  {(B)} The equations for a CTLN network.
{(C)} (Left) An oriented graph on 3 nodes, with the corresponding adjacency matrix below.  (Right) Network activity follows the arrows in the graph, with peak activity occurring sequentially in the cyclic order 123.  {(D)} Cliques correspond to stable fixed points, but only if they are {\it target-free} cliques \rkatie{(see Theorem~\ref{thm:Thm2})}.  The clique $12$ is target-free, but $23$ is not.  Unless otherwise noted, all simulations have parameters $\varepsilon=0.25$, $\delta=0.5$, and $\theta=1$.}
\label{fig:CTLN-model}
\end{center}
\vspace{-.2in}
\end{figure}

We refer to threshold-linear networks of the form~\eqref{eq:network}, with $b_i=\theta>0$ and $W = W(G,\varepsilon,\delta)$ as in~\eqref{eq:binary-synapse}, as the {\it combinatorial threshold-linear network} (CTLN) model.  Note that this model is completely specified by the choice of directed graph, $G$, along with three positive real parameters: $\varepsilon, \delta,$ and $\theta$ (see Figure~\ref{fig:CTLN-model}A-C). 

\rkatie{In order to ensure that the graph $G$ is most meaningful for the dynamics, it is useful to require that any pair of nodes $i, j$ with a single unidirectional edge $i \to j$ (but $j\not\to i$) cannot support a stable fixed point, so that activity will flow along directed edges. This is guaranteed if the $2\times 2$ matrix $(-I+W)_{\{i,j\}}$ for the directed edge $ i \to j$ is always unstable. Specifically, the eigenvalues of the following matrix, 
$$(-I+W)_{\{i,j\}} = \left( \begin{array}{cc} -1 &  -1 - \delta\\ -1 + \varepsilon & -1\end{array} \right),$$
should not both have strictly negative real part.
Since the trace is negative, this occurs precisely when $\det (-I+W)_{\{i,j\}} = \varepsilon(1+\delta) - \delta <0$.  We thus define the {\it legal range} for a CTLN as follows.
\begin{definition}
We say that the CTLN parameters $\varepsilon, \delta, \theta$ are in the \emph{legal range} if  $\varepsilon, \delta, \theta > 0$ and $\varepsilon < \delta/(\delta + 1)$.
\end{definition}
}

\rkatie{We interpret a CTLN as modeling a network of $n$ excitatory neurons in the presence of strong background inhibition  (Figure~\ref{fig:CTLN-model}A). When $j \not\to i$, we say $j$ \emph{strongly inhibits} $i$; when $j \to i$, we say $j$ \emph{weakly inhibits} $i$. The strong inhibition is just the global background inhibition; while the weak inhibition can be thought of as the sum of an excitatory connection and global inhibition.  
%Note that because $-1-\delta < -1 < -1+\varepsilon$, when $j \not\to i$, neuron $j$ inhibits $i$ \emph{more} than it inhibits itself via its leak term; when $j \to i$, neuron $j$ inhibits $i$ \emph{less} than it inhibits itself.  
These differences in inhibition promote the flow of activity along the arrows of the graph. 
}

We have seen that competitive TLNs have bounded activity that remains confined to a box, $\prod_{i=1}^n [0,b_i]$ (Lemma~\ref{lemma:bounded}). For CTLNs, the corresponding box is $[0,\theta]^n$.  
What can we say about the dynamics inside this box?  
In the simplest cases, where all interactions are equal so that $W_{ij} = w$ for all $i \neq j$, we have two extremes: $w = -1-\delta$ or $w = -1+\varepsilon$.  If inhibition is strong, so that $w = -1-\delta < -1$, we obtain a classical winner-take-all (WTA) network.  Such a network has $n$ stable fixed points, one corresponding to each node, and no other attractors.  The activity always converges to the fixed point of the ``winning'' neuron.  If, on the other hand, inhibition is weak, so that $w = -1+\varepsilon > -1$, then the network synchronizes and activity always converges to a single stable fixed point in which all nodes have equal activity. Neither case is particularly exciting.  

When a competitive network has {\it both} strong and weak inhibition, things get more interesting, particularly when the interactions are asymmetric. Such networks can exhibit a variety of stable and unstable fixed points, and dynamic attractors appear to correspond to a key subset of these unstable fixed points \cite{core-motifs}.  Thus, we are motivated to understand the full collection of fixed points of a CTLN (or a TLN), not only those that are stable.  Moreover, we are particularly interested in networks with \underline{no} stable fixed points, as these will be guaranteed to have dynamic attractors.

\section{Mathematical results}\label{sec:math-results}

We begin with some results that hold for the more general class of nondegenerate competitive TLNs. \rkatie{Theorem~\ref{thm:parity} is an index theorem for the fixed points of a TLN, which has a number of important consequences.  In particular, it implies that the total number of fixed points is always odd and provides an upper bound on the number of stable fixed points.  Theorem~\ref{thm:Thm1} gives conditions on the connectivity matrix $W$ that guarantee that the TLN has \underline{no} stable fixed points.  Specializing to CTLNs, we obtain Theorem~\ref{thm:Thm1b}, which tells us that oriented graphs with no sinks yield networks with no stable fixed points.  At the other extreme, Theorem~\ref{thm:Thm2} shows that \emph{cliques} correspond to stable fixed points (provided that the embedding is \emph{target-free}). Moreover, we conjecture that these are the only subgraphs that correspond to stable fixed points.  These results are illustrated with numerous examples in Section~\ref{sec:simulations}.} 
The proofs are postponed to Section~\ref{sec:proofs}.

\subsection{Index and parity} 
For each fixed point of a competitive TLN $(W,b)$, labeled by its support $\sigma \in \FP(W,b)$, we define the \emph{index} as
$$\idx(\sigma) \od \sgn \det(I-W_\sigma).$$
Since we assume our TLNs are nondegenerate, we have $\det(I-W_\sigma) \neq 0$ and thus $\idx(\sigma) \in \{\pm 1\}$.  Note that if $\sigma$ is the support of a \emph{stable} fixed point, then the eigenvalues of $-I+W_\sigma$ must all have negative real part, and so those of $I-W_\sigma$ all have positive real part. This implies that $\idx(\sigma) = +1$ for all stable fixed points.

The following theorem indicates that fixed points with index $+1$ and $-1$ are almost perfectly balanced. It also tells us that the parity of the total number of fixed points is always odd.

\begin{theorem}[parity]\label{thm:parity}
Let $(W,b)$ be a competitive nondegenerate TLN on $n$ nodes, with $b_i>0$ for all $i \in [n]$. Then
$$\sum_{\sigma \in \FP(W,b)} \idx(\sigma) = + 1.$$
In particular, the total number of fixed points $|\FP(W,b)|$ is always odd.
\end{theorem}

As an immediate corollary, we obtain an upper bound on the number of stable fixed points.  Note that $|\FP(W,b)| \leq 2^n -1$, \rkatie{since nondegenerate TLNs have at most one fixed point per support (Corollary~\ref{cor:fp-bound}) and the origin (empty support) is never a fixed point.} Because stable fixed points all have index $+1$, their upper bound is at most half of $2^n-1$.

\begin{corollary}
The number of stable fixed points of a competitive nondegenerate TLN on $n$ nodes, \rkatie{with $b_i>0$ for all $i \in [n]$,} is at most $2^{n-1}$.
\end{corollary}

\subsection{Competitive networks with no stable fixed points}
Historically, much of the mathematical theory of TLNs has been developed in the symmetric case with a focus on networks with guaranteed convergence to stable fixed points \cite{HahnSeungSlotine, XieHahnSeung, Hahn2000, flex-memory, net-encoding, pattern-completion}.  These networks can serve as models of associative memory storage and retrieval, similar to Hopfield networks, with static memory patterns encoded as stable fixed points.  But the brain also exhibits {\it dynamic} memory patterns, such as rhythms and sequences. This type of activity is more naturally modeled with dynamic attractors, such as limit cycles and higher-dimensional attractors, not stable fixed points. Can we find conditions that guarantee that a TLN only exhibits dynamic attractors?

Our next result establishes sufficient conditions such that a competitive network has no stable fixed points (i.e., no steady states).  The statement of the theorem makes use of a simple\footnote{A graph is {\it simple} if it has no multiple edges and no self-loops.  Simple graphs have binary adjacency matrices with zeros on the diagonal.}
directed graph $G_W$ on $n$ vertices, that is defined from the $n \times n$ connectivity matrix $W$ as follows:  
\begin{equation}\label{eq:G_W}
G_W \text{ has an edge from } j \rightarrow i \;\Leftrightarrow \;W_{ij}> -1 \;\; (\text{for } i \neq j).
\end{equation}
Note that $W_{ij}$ represents the influence of node $j$ on node $i$.  The edges of $G_W$ correspond to inhibitory interactions that are weaker than the self-inhibition of each node. If $W$ is the matrix of a CTLN with graph $G$, then $G_W = G$.

The next theorem uses the following graph-theoretic terminology: a graph is {\it oriented} if there are no bidirectional edges, and a {\it sink} is a vertex with no outgoing edges.  The proof is given in Section~\ref{sec:proofs}. 

\begin{theorem}\label{thm:Thm1}
Consider a competitive \rkatie{nondegenerate} threshold-linear network with connectivity matrix $W$, associated graph $G_W$, and uniform inputs $b_i=\theta$.  Suppose that:
\begin{itemize}
\item[(i)] $G_W$ is an oriented graph with no sinks, and
\item[(ii)] whenever $j \rightarrow i$ in $G_W$, $W_{ij} < \dfrac{1}{W_{ji}}$.
\end{itemize}
Then the network~\eqref{eq:network} has no stable fixed points.  Moreover, the network activity is bounded.
\end{theorem}

In the case of a CTLN, it is easy to see that condition (ii) of Theorem~\ref{thm:Thm1} is always satisfied, provided the parameters are within the {\it legal range}, so that $\varepsilon < \delta/(1+\delta)$. In fact, this was one of our motivations for the definition of legal range. We thus have the following result, obtained by specializing Theorem~\ref{thm:Thm1} to the CTLN case.

\begin{theorem}\label{thm:Thm1b}
Let $G$ be an oriented graph with no sinks, and consider an associated  \rkatie{nondegenerate} CTLN with $W = W(G,\varepsilon,\delta)$ for $\varepsilon$ and $\delta$ in the legal range.  Then the network has bounded activity and no stable fixed points.
\end{theorem}
  
Figure~\ref{fig:CTLN-model}C displays the smallest oriented graph with no sinks, together with the activity of the corresponding CTLN model.  The solutions to~\eqref{eq:network} for this $W$ always converge to the same perfectly periodic trajectory, irrespective of the initial conditions. It is surprisingly nontrivial to explicitly prove that this limit cycle exists.  A recent proof was given in \cite{Horacio-paper} for the existence of a unique limit cycle in CTLNs whose graph is a 3-cycle, as in Figure~\ref{fig:CTLN-model}C, as well as a $k$-cycle more generally.

\subsection{Stable fixed points in the CTLN model}

Theorem~\ref{thm:Thm1b} gave conditions on the graph $G$ of a CTLN that guaranteed the absence of stable fixed points. In this section, we present a theorem and a conjecture relating certain graph structures to the existence of stable fixed points.

To state the next theorem, we need a few graph-theoretic definitions.
A subset of vertices $\sigma$ is a {\em clique} of $G$ if the nodes in $\sigma$ are all-to-all bidirectionally connected, i.e. $i \leftrightarrow j$ for all pairs $i,j \in \sigma$.  We say that a vertex $k$ is a {\it target} of $\sigma$ if $k \notin \sigma$ and $i \rightarrow k$ for each $i \in \sigma$.  If the clique $\sigma$ has no targets, we say that it is {\it target-free}.  A clique is {\it maximal} if it is not contained in any larger clique of $G$.  Note that all target-free cliques are necessarily maximal, but maximal cliques need not be target-free.  For example, the graph in Figure~\ref{fig:CTLN-model}D has two maximal cliques, but only one of them is target-free.  It turns out that the only cliques that can support stable fixed points are target-free cliques.

\begin{theorem}\label{thm:Thm2}
Let $G$ be a simple directed graph, and consider an associated  \rkatie{nondegenerate} CTLN with $W = W(G,\varepsilon, \delta)$ for any choice of the parameters $\varepsilon, \delta, \theta >0$ with $\varepsilon < 1.$
If $\sigma$ is a clique of $G$, then there exists a stable fixed point with support $\sigma$ if and only if $\sigma$ is target-free.
\end{theorem}

\noindent The proof is given in Section~\ref{sec:proofs}.  Note that the result is valid for parameters beyond the legal range, including $\delta/(\delta+1) \leq \varepsilon <1$.

While Theorem~\ref{thm:Thm2} identifies precisely which cliques support stable fixed points, there may still exist additional stable fixed points that do not correspond to cliques.  We have not, however, been able to find any such example.  This leads us to the following conjecture:

\begin{conjecture}\label{conj} Consider a \rkatie{nondegenerate} CTLN model $W = W(G,\varepsilon, \delta)$, where 
$G$ is a simple directed graph \rkatie{and $\varepsilon$ and $\delta$ are within the legal range}.  There exists a stable fixed point with support $\sigma$ 
if and only if $\sigma$ is a target-free clique of $G$.
\end{conjecture}

The conjecture was previously proven to hold in the case of symmetric $W$, i.e.\ when all the edges in $G$ are bidirectional \cite{pattern-completion}.  Note that in that context, a clique is target-free precisely when it is maximal.  Outside of the symmetric case, the conjecture has been proven in a few other settings, such as for graphs on $n \leq 4$ nodes or within a particular region of the legal range of parameters $\varepsilon$ and $\delta$.  These results and all other current evidence in support of the conjecture are collected in \cite{stable-fp-paper}.

\section{Simulations}\label{sec:simulations}

\rkatie{The CTLN model captures a surprisingly rich diversity of nonlinear dynamics.   In this section, we provide a variety of examples to illustrate both this diversity and the theorems from the previous section.  Unless otherwise noted, all simulations have parameters $\varepsilon=0.25$, $\delta=0.5$, and $\theta=1$.  These are referred to as the \emph{standard parameters}.  When the parameters are identical, differences in network dynamics arise solely as a result of differences in the underlying graph $G$.} 

\rkatie{The Matlab package CTLN Basic 2.0 can be used to reproduce the simulations in Figures~\ref{fig:CTLN-model}--\ref{fig:phone-number}.  In particular, graphs and initial conditions are provided for each of the figures, with the exception of Figure~\ref{fig:state-transition}.  The package is available at \url{https://github.com/nebneuron/CTLN-Basic-2.0}. }

\subsection{Dynamic diversity from network connectivity}\label{sec:diversity}

\begin{figure}[!h]
\begin{center}
%\vspace{.1in}
\includegraphics[width=5in]{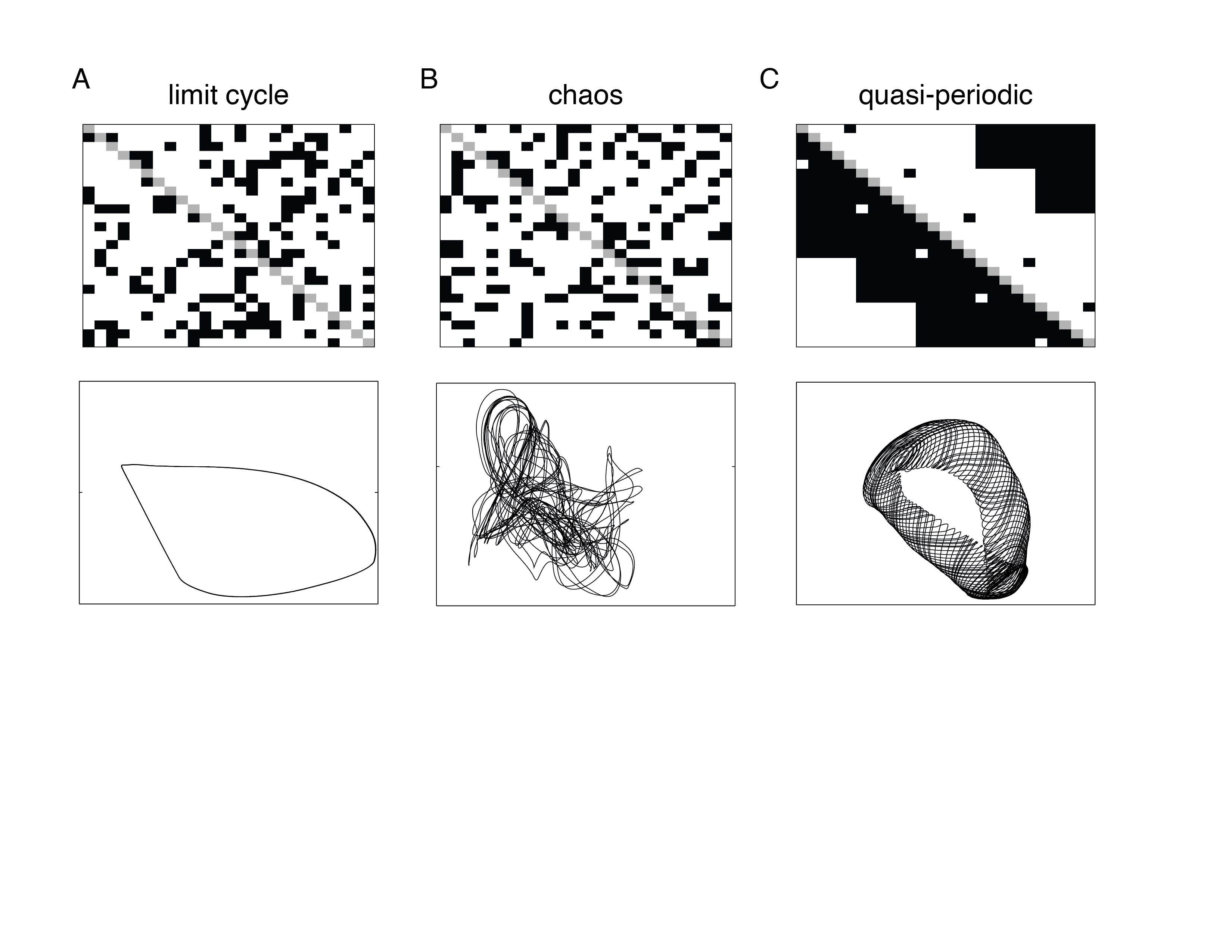}
\vspace{-1.2in}
\caption{{\bf Nonlinear dynamics of the CTLN model.}  Top panels show adjacency matrices for three oriented graphs on $n=25$ nodes with no sinks, satisfying the conditions of Theorem~\ref{thm:Thm1b}.  (Black = 1, white = 0, and gray = diagonal elements, which are ignored.)  The corresponding CTLN model networks produce (A) a limit cycle, (B) a chaotic attractor, and (C) quasi-periodic behavior.  Bottom panels show random two-dimensional projections of the 25-dimensional trajectories. }
\label{fig:chaos}
\end{center}
\vspace{-.3in}
\end{figure}

Figure~\ref{fig:chaos} displays adjacency matrices for three different graphs on $n=25$ nodes, along with two-dimensional projections of solutions that are periodic (A), chaotic (B), and quasi-periodic (C).  \rkatie{Since each of these graphs is oriented with no sinks, Theorem~\ref{thm:Thm1b} guarantees that the networks have bounded activity but no stable fixed points.  Note, however, the large variety of dynamics that can arise by varying the graph structure.} For example, the quasi-periodic behavior in Figure~\ref{fig:chaos}C is shaped by a highly structured graph; and this graph differs markedly from the ones in Figures~\ref{fig:chaos}A,B.

What aspects of the connectivity graph $G$ determine the emergent dynamics of the network?  
Are local properties of the connectivity matrix sufficient to predict the behavior that will emerge?  One possibility is that the degree profile, that is the list of in-degrees and out-degrees of each node, is predictive.

\begin{figure}[!hb]
\begin{center}
%\vspace{-.1in}
\includegraphics[width=6.25in]{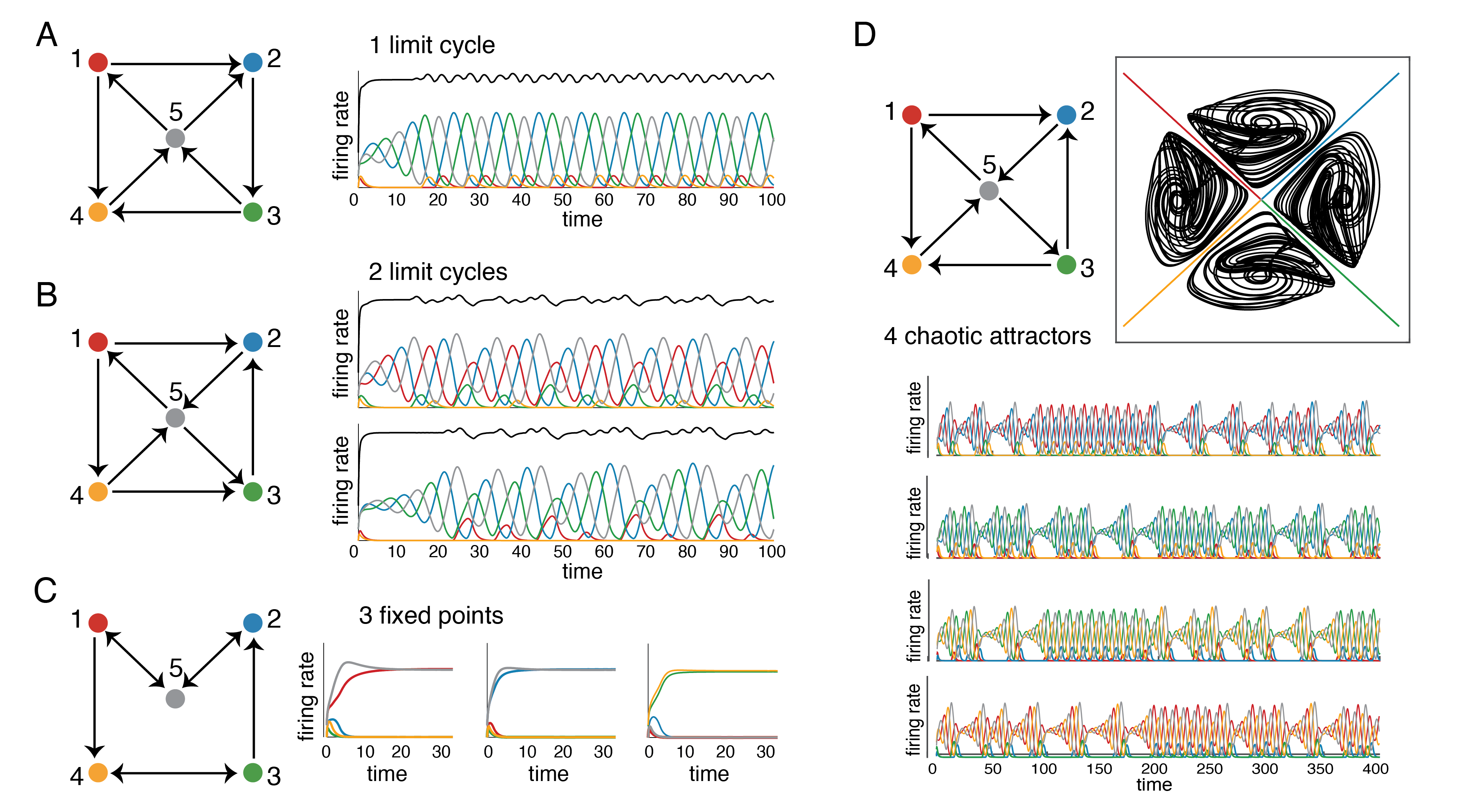}
\caption{{\bf Degree-matched networks with different dynamics.}  (A--D) Four graphs are shown together with the attractors that emerge from their corresponding CTLNs. Each activity plot shows a solution arising from a different initial condition, where the colors match the nodes in the graph. Networks A-D have precisely the same degree profile, but significantly different dynamics: limit cycles (A, B), stable fixed points (C), and strange/chaotic attractors (D).  A two-dimensional projection of the four strange/chaotic attractors is shown in D. 
}
\label{fig:deg-matched}
\end{center}
%\vspace{-.1in}
\end{figure}

Figure~\ref{fig:deg-matched} makes it clear that the degree profile is insufficient to predict the patterns of activity that emerge.  The four networks shown in panels A-D have graphs on $n=5$ nodes with exactly the same degree profile: $\{(2,2),(2,1),(2,1),(1,2),(1,2)\}$, where $(a,b)$ indicates in-degree $a$ and out-degree $b$ of a single node. 
Despite having identical local properties, dramatically different dynamics arise: limit cycles in networks A and B, stable fixed points in network C, and strange/chaotic attractors in network D.  These CTLNs thus exhibit {\it emergent dynamics} in the strongest sense, as differences in activity depend only on the global structure of the network, and not on local properties of individual nodes (which are identical across networks A-D).
 Note that the graphs in panels A, B, and D are all oriented, and so the absence of stable fixed points is guaranteed by Theorem~\ref{thm:Thm1b}.  In contrast, the graph in C is not oriented; its stable fixed points are fully predicted by Theorem~\ref{thm:Thm2}.

The networks in Figure~\ref{fig:deg-matched}B--D also exhibit multiple coexisting attractors, with different initial conditions leading to different patterns of activity.  For example, network B has two distinct limit cycles, each of which can be accessed by changing initial conditions -- without changing any network parameters. Similarly, network C has three stable fixed points, and network D has four strange/chaotic attractors.
While each of these networks possesses multiple attractors, all attractors within a given network are of the same type.  Is it possible for different types of attractors to coexist within a single network?  

\begin{figure}[!h]
\begin{center}
\vspace{-.1in}
\includegraphics[width=5.5in]{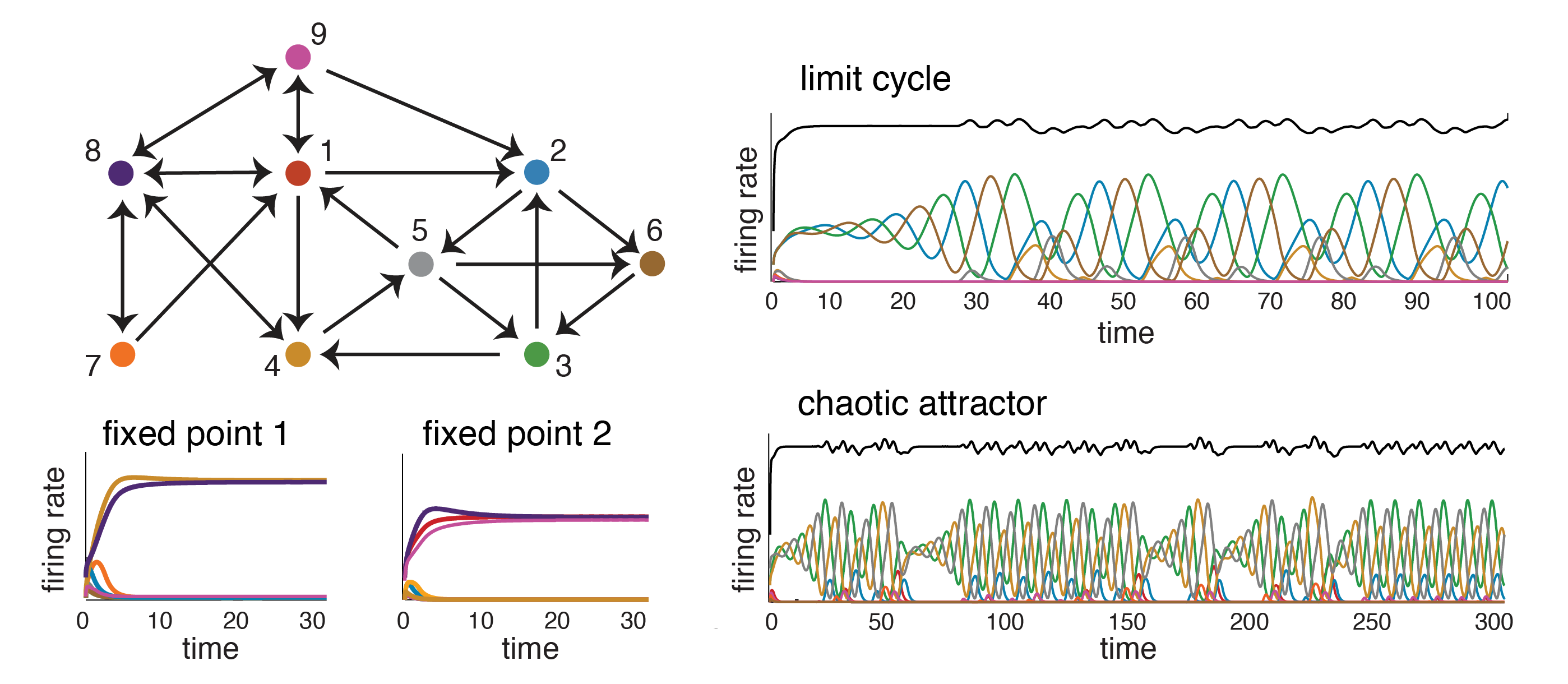}
\caption{{\bf Variety of emergent dynamics in a single network with $n=9$ nodes.}  Solutions corresponding to different initial conditions are shown.  The network has four attractors: two stable fixed points, one limit cycle, and one chaotic attractor.  The equations for the dynamics are identical in each case; only the initial conditions differ.}
\label{fig:coexistence}
\end{center}
\vspace{-.2in}
\end{figure}

The network in Figure~\ref{fig:coexistence} shows that 
different types of attractors can coexist: two stable fixed \rkatie{points}, a limit cycle, and a chaotic attractor.  Once again, the selected attractor depends only on the choice of initial conditions, without any change in network parameters. And the stable fixed points can again be predicted using Theorem~\ref{thm:Thm2}.

\subsection{Emergent sequences}

In Figure~\ref{fig:CTLN-model}C, Figure~\ref{fig:deg-matched}A--B, and Figure~\ref{fig:coexistence}, we saw limit cycles in which nodes were activated in a regular sequence.  Sequential patterns of activity are in fact quite common in CTLNs, as competition between nodes results in a tendency for neurons to ``take turns'' in reaching their peak firing.
The emergent sequences, however, are often irregular and surprising.  
\begin{figure}[!h]
\begin{center}
\includegraphics[width=5.75in]{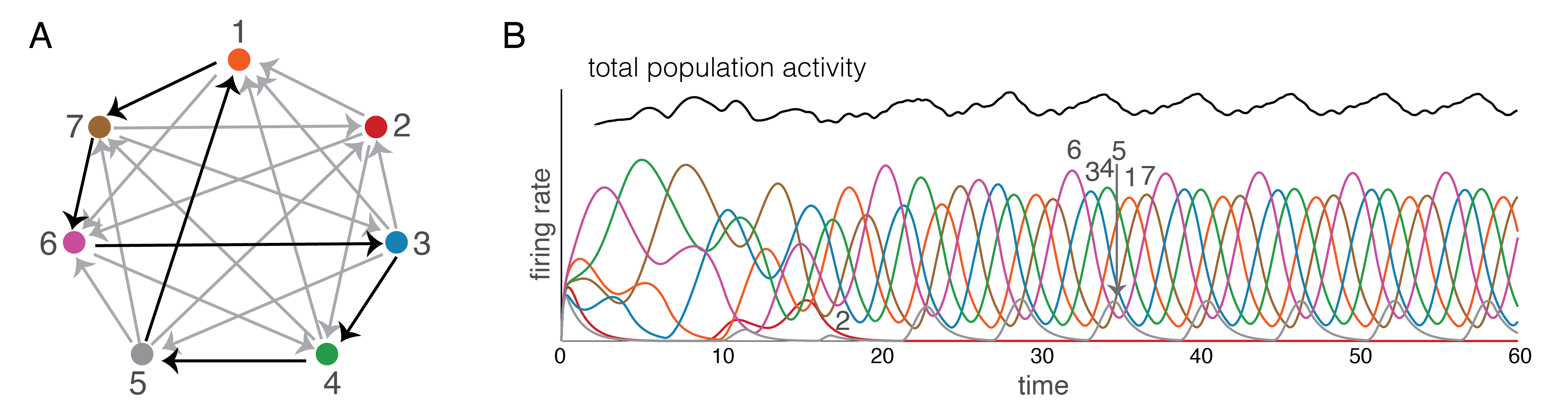}
\vspace{.1in}
\caption{{\bf Emergence of an irregular sequence.} {(A)} A graph on 7 nodes with a particular cycle highlighted in black.  The CTLN model makes no distinction between black and gray edges, but black edges are highlighted here because they correspond to the emergent sequence of activation.   {(B)} Node 2 decays to zero after a short period of transient activity, while the remaining nodes settle into a limit cycle of sequential activation with ordering 634517.  The same sequence emerges irrespective of initial conditions, and is robust to small perturbations of the matrix $W$.}
\label{fig:graph2sequence}
\end{center}
\vspace{-.1in}
\end{figure}
Figure~\ref{fig:graph2sequence} displays a network with $n=7$ nodes and a single emergent sequence (irrespective of initial conditions). Although the graph has many cycles, the sequential activity follows only one cycle in the graph.  Moreover, the activity of node 2 always decays to zero, even though there are other nodes (3 and 5) with smaller in-degree whose activity persists in the attractor. This provides another example where local properties of the graph are not sufficient to predict emergent dynamics; the resulting sequence has been shaped by the structure of the graph as a whole.   

\begin{figure}[!h]
\begin{center}
\includegraphics[width=6.5in]{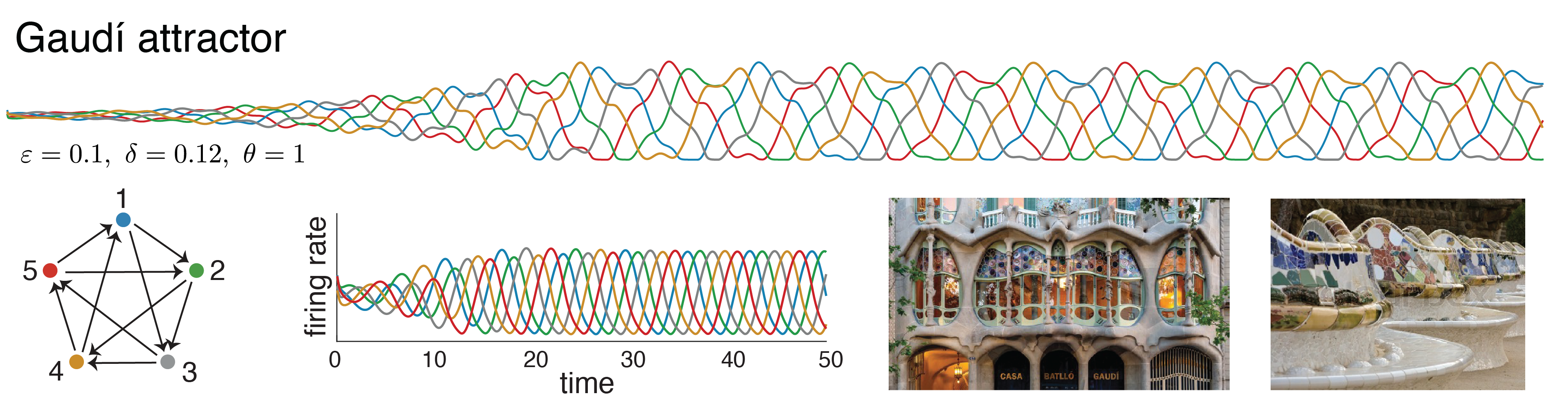}
\caption{{\bf The Gaudi attractor.} A CTLN for a cyclically symmetric tournament on $n=5$ nodes produces two distinct attractors, depending on initial conditions.  We call the top one the Gaudi attractor because the undulating curves are reminiscent of the architect from Barcelona.  Note the different parameters for this CTLN: $\varepsilon=0.1$ and $\delta=0.12$. % For Gaudi attractor, e=.1, d=.12, X0 = [0.2160   0.2400    0.2277    0.2259    0.2443];
}
\label{fig:Gaudi}
\end{center}
\vspace{-.1in}
\end{figure}

\FloatBarrier
Figure~\ref{fig:Gaudi} shows the dynamic attractors of a CTLN whose graph is a cyclically symmetric tournament on
$n = 5$ nodes. For some initial conditions, the dynamics converge to a somewhat boring limit cycle with the firing
rates $x_1(t),\ldots , x_5(t)$ all peaking in the expected sequence, 12345 (bottom middle). However, for a different
set of initial conditions, the solution converges to the beautiful but unusual attractor displayed at the
top.  

\begin{figure}[!h]
\begin{center}
\includegraphics[width=7in]{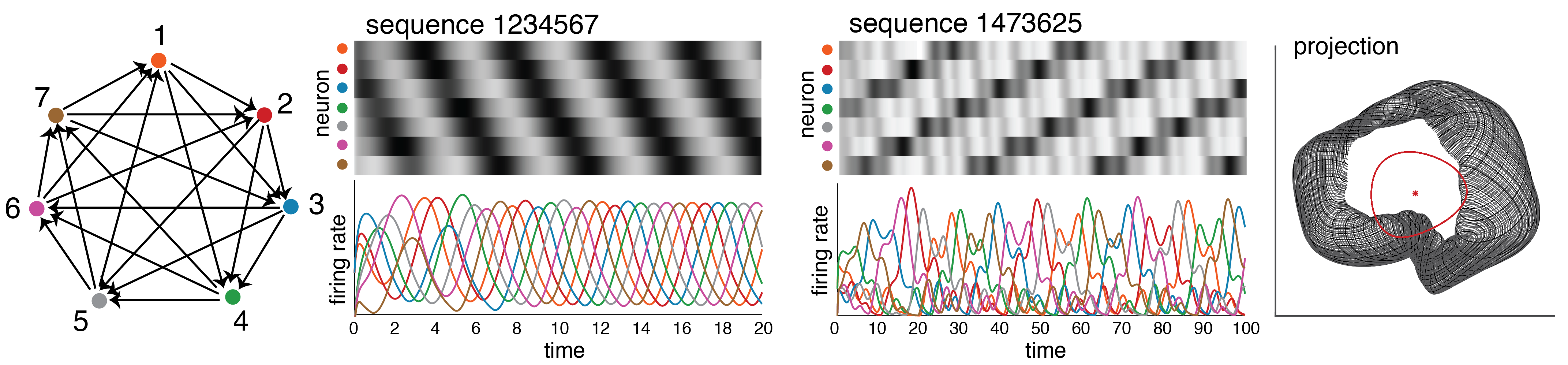}
\caption{{\bf A cyclically symmetric CTLN on $\mathbf{n=7}$ nodes.} {(Left)} The graph is a cyclically symmetric tournament on $n=7$ nodes.  {(Middle)} The corresponding CTLN has two attractors: one limit cycle with sequence 1234567, and one quasiperiodic attractor whose sequence corresponds to different cycle in the graph. {(Right)} A $2$-dimensional projection of the two attractors and the unique (unstable) fixed point of the network. The limit cycle and fixed point are displayed in red, while the quasiperiodic attractor is the torus-like trajectory shown in black.}
\label{fig:7star}
\end{center}
\vspace{-.2in}
\end{figure}

Similar behavior emerges from a CTLN whose graph is a generalization of this cyclic tournament structure to $n = 7$ nodes (see Figure~\ref{fig:7star}).  For some initial conditions, a simple limit cycle emerges with the expected sequence of peaks 1234567 (middle left).  For different initial conditions, another attractor emerges (middle right), which is quasiperiodic with a torus-like trajectory shown on the right.  Interestingly, this CTLN has a unique fixed point, which is unstable and has all nodes active, and both of these attractors are accessible from perturbations of that fixed point.  This same phenomenon holds for the CTLN in Figure~\ref{fig:Gaudi} as well.  

\rkatie{The dynamics in Figure~\ref{fig:7star} are reminiscent of traveling wave solutions in bump attractor networks \cite{ErmentroutTerman2010, Burak2012FundamentalLO, JN2011, ItskovTsodyks2011}. The traveling wave interpretation is natural if we consider the nodes as spatially organized as in a ring model.  In classic ring models, the connectivity strength typically drops off with distance and the connectivity matrix is symmetric.  In contrast, the network in Figure~\ref{fig:7star} only has ``forward edges" around the cycle, and is thus far from symmetric. Nevertheless, due to the cyclic symmetry, one observes a bump of activity traveling around the cycle. Interestingly, the second attractor of this network has a different sequence of activity that does not respect the same geometric organization. In this solution, the activity jumps between nodes that are neighbors in the graph, but not neighbors in the ring. However, using a different geometric arrangement of the nodes, so that ordering on the ring is 1473625, the activity may again be thought of as a traveling wave.}

\subsection{Complex rhythms}

\begin{figure}[!h]
\begin{center}
\includegraphics[width=5.5in]{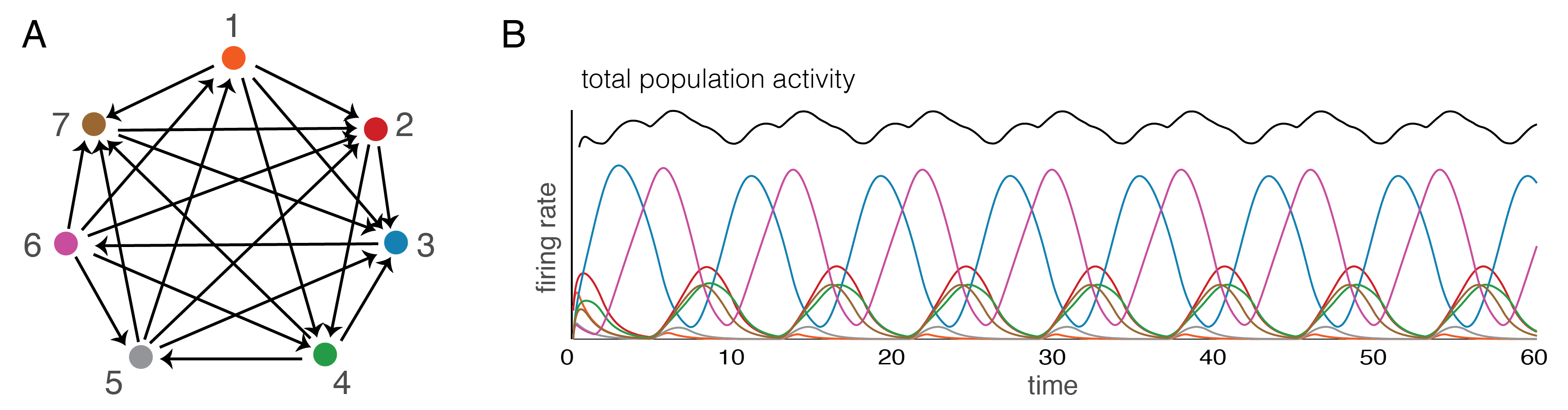}
\caption{{\bf Emergence of a complex rhythm.}  A graph on 7 nodes (left) yields a CTLN model whose activity always settles into the same limit cycle (right).  The activity in this limit cycle is a rhythmic, with some nodes that are quasi-synchronous rather than sequential in their activation.}
\label{fig:purring}
\end{center}
\vspace{-.2in}
\end{figure}

Limit cycles need not be perfectly sequential -- they can also display complex rhythms, including synchronous or quasi-synchronous activity for a subset of the nodes.  The network in Figure~\ref{fig:purring} has two high-activity nodes (3 and 6) that peak at different times, while nodes 2, 4 and 7 are approximately synchronous.  Note that node 6 has the highest peak activity, even though it has the lowest in-degree among all nodes in the network.

\begin{figure}[!h]
\begin{center}
\includegraphics[width=3.8in]{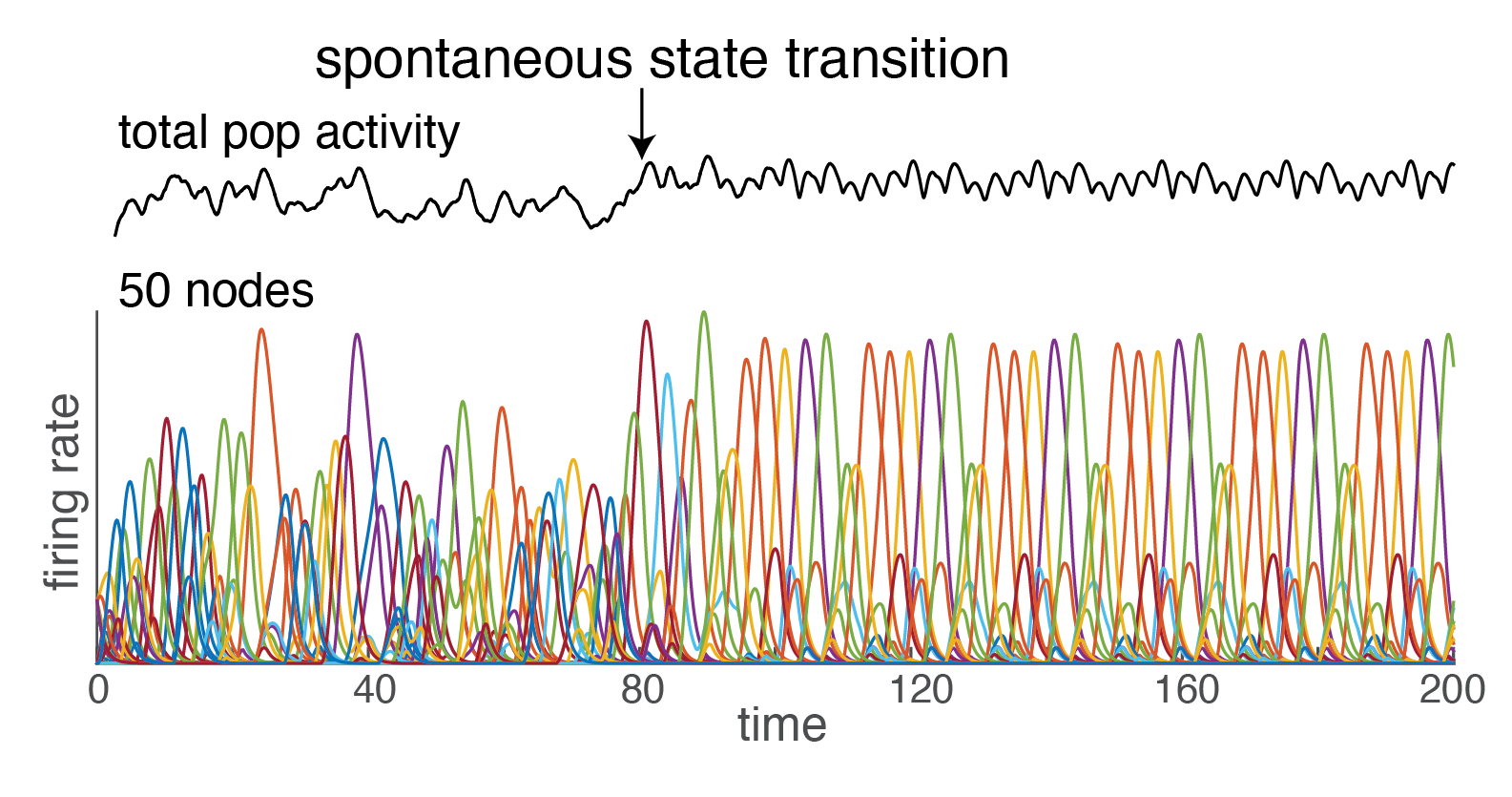}
\caption{{\bf Spontaneous state transition.}  A random network of $50$ nodes satisfying the conditions of Theorem~\ref{thm:Thm1b} exhibits a spontaneous state transition from irregular to periodic behavior.  Total population activity (black trace) is the sum of the activity for all 50 nodes.}
\label{fig:state-transition}
\end{center}
\end{figure}

Using Theorem~\ref{thm:Thm1b}, we can also generate large random networks that are guaranteed never to settle into a steady state, and thus possess interesting dynamic attractors.  Such networks can exhibit spontaneous transitions between distinct patterns of network activity.  In Figure~\ref{fig:state-transition}, the total population activity trace has a sharp qualitative change around $t=80$; this is reminiscent of state transitions in cortical networks observed during light anesthesia and sleep \cite{JN2009}.

\rkatie{The simulations in Figures~\ref{fig:CTLN-model}--\ref{fig:state-transition} yield attractor types that are not exclusive to CTLNs; similar behaviors have been observed in many other nonlinear systems.  What is remarkable is that they can all be produced within the same model family, and with the same parameters, by varying only a simple graph of connectivity.  Moreover, this model family (TLNs/CTLNs) is surprisingly mathematically tractable.  Indeed, it is this tractability that enabled the discovery of such a variety of dynamics without computationally intensive searches of the TLN parameter space. For example, the chaotic attractors in Figure~\ref{fig:deg-matched}D are reminiscent of the R\"ossler attractor, but were naturally discovered by exploring small oriented graphs with no sinks.  The Gaudi attractor in Figure~\ref{fig:Gaudi} was also discovered in this way, and appears to be novel.}

\FloatBarrier
\subsection{Network engineering}\label{sec:engineering}

The power of  Theorems~\ref{thm:parity},~\ref{thm:Thm1b} and~\ref{thm:Thm2} is that they enable us to reason mathematically about the graph $G$ in order to make strong yet accurate predictions about the resulting network dynamics.  Such mathematical results are thus valuable tools for designing networks with prescribed dynamic properties.

\begin{figure}[!h]
\begin{center}
\includegraphics[width=6.75in]{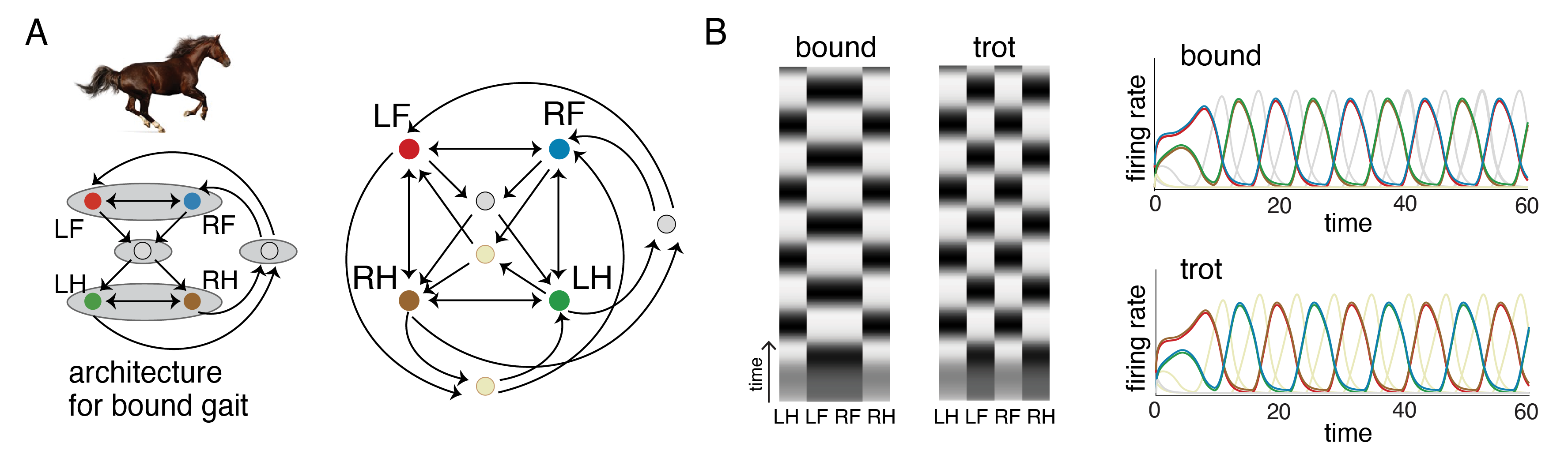}
\vspace{.1in}
\caption{{\bf A Central Pattern Generator circuit for quadruped motion.} {(A)} (Left) A \emph{cyclic union} architecture on 6 nodes that produces the `bound' gait.  (Right) The graph on 8 nodes is formed from gluing together architectures for the individual gaits, `bound' and `trot'.  Note that the positions of the two hind legs (LH, RH) are flipped for ease of drawing the graph. {(B)} The network produces limit cycles corresponding to two distinct gaits. Convergence to one or the other is determined only by initial conditions.}
\label{fig:quadruped}
\vspace{-.15in}
\end{center}
\end{figure}

Rhythmic patterns of activity, supporting locomotion and other functions, arise in Central Pattern Generator circuits (CPGs) throughout the nervous system \cite{Marder-CPG, Yuste-CPG}.  The CTLN model provides a natural framework for CPGs.  For example, Figure~\ref{fig:quadruped} shows that limit cycles corresponding to two different quadruped gaits, `bound' (similar to gallop) and `trot', can coexist in the same network, with the network selecting one pattern over the other based solely on initial conditions. 

The network in Figure~\ref{fig:quadruped} was produced by essentially ``gluing together" two architectures that would produce the desired gaits, identifying the graphs along the nodes corresponding to each leg.  The left panel in A shows the isolated graph that produces the single gait, `bound'.   Notice that the clique between the left and right front legs (LF and RF) ensures that those nodes co-fire; the same is true for the clique between the left and right hind legs (LH and RH).  But since each of these cliques has a target, neither of them can a support stable fixed point (by Theorem~\ref{thm:Thm2}).  The activity simply flows forward from the clique to one of the (gray) auxiliary nodes, and then to the next clique in the cycle.  A similar network was created for the `trot' gait, with appropriate pairs of legs joined by a clique. The larger graph in Figure~\ref{fig:quadruped}A is the result of gluing together the smaller graphs for these two gaits.

\begin{figure}[!h]
\begin{center}
\vspace{-.05in}
\includegraphics[width=6.75in]{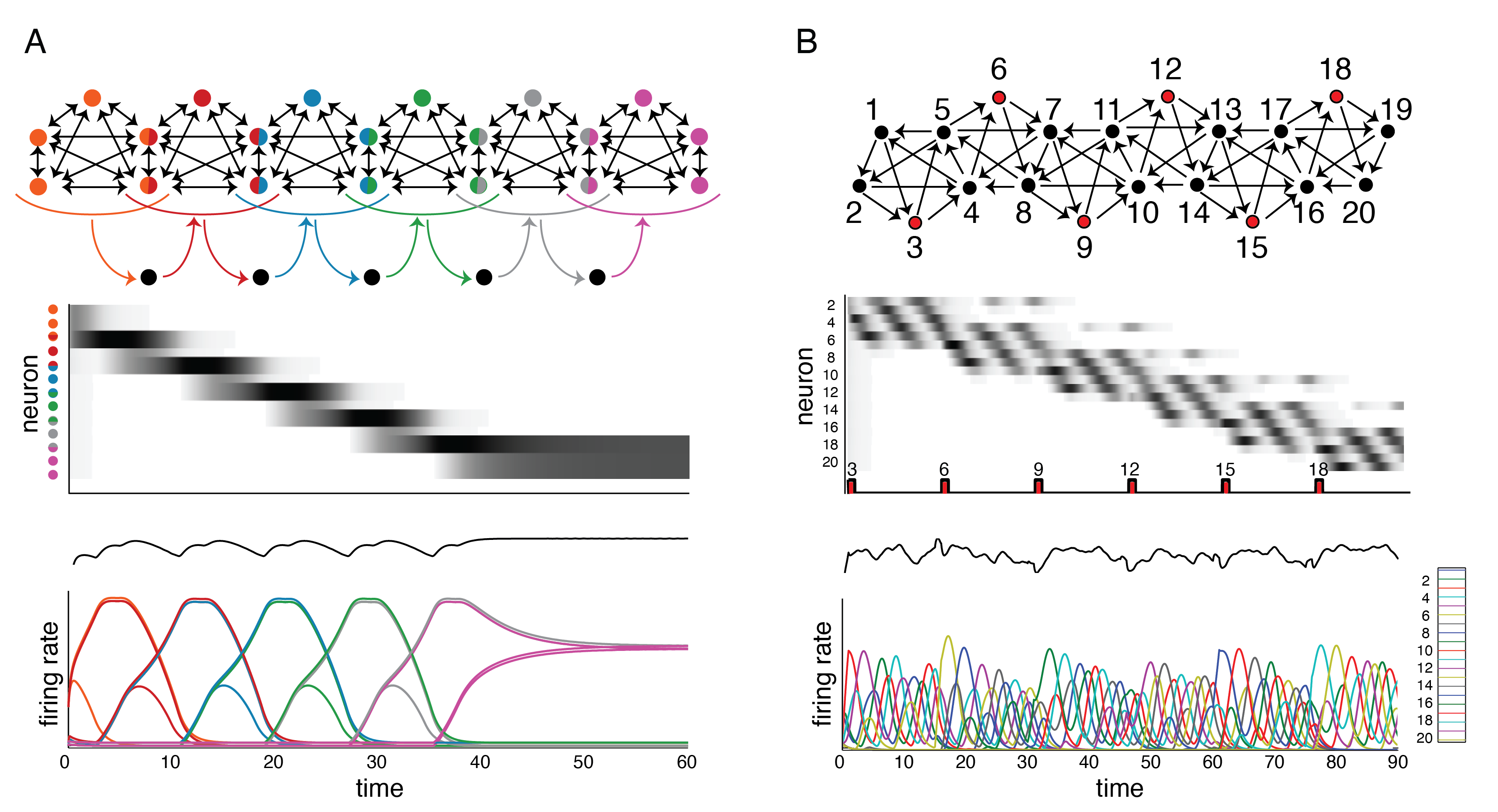}
\vspace{-.05in}
\end{center}
\caption{{\bf Regular and irregular sequential activity from modular architectures.}  {(A)} (Top) A chain of six overlapping 5-clique modules, each in a different color.  Nodes belonging to two adjacent 5-cliques are double-colored.  Black nodes receive output edges from all nodes in one 5-clique, and feed forward onto all nodes in the next 5-clique.  Note that only the final 5-clique (purple) is target-free.  (Middle) The solution of the network when the nodes in the first 5-clique are initialized to $0.1$, and all other nodes are initialized to 0. Darker regions correspond to higher firing rates; note that the outer target nodes are omitted.  (Bottom) Same solution as in middle plot.  The firing rate curves are colored by the nodes in the graph, with overlap nodes receiving both colors.  The activity moves slowly from one clique to the next until it stabilizes on the last clique, which is target-free.  The total population activity is given by the black trace above.   {(B)} (Top) A chain of six overlapping modules, where each module is a cyclic tournament on five nodes (the same graph as in Figure~\ref{fig:Gaudi}).  Nodes highlighted in red receive small kicks during the simulation.  (Middle) The solution of the network when nodes 1 and 2 are initialized to $0.1,$ and all other nodes are initialized to 0.  Initially the network activity is confined to the first module, and cycles among those nodes.  Every 15 time units, a small kick is given to the middle node in the next module.  The timing of these kicks and the label of the affected node are shown as red pulses along the bottom of the plot.  (Bottom) Same solution as in the middle plot, with individual firing rates shown in color.  
}
\label{fig:modular}
\end{figure}

One key to the construction in Figure~\ref{fig:quadruped} was the use of cliques with targets to ensure a desired flow of activity. More generally, Theorem~\ref{thm:Thm2} enables us to construct networks that continually transition between cliques without getting ``stuck,'' since we know precisely which cliques in a graph correspond to steady states. Figure~\ref{fig:modular}A depicts a network with a series of overlapping cliques, only the last of which is target-free.  We can also engineer networks by patching together modules that individually yield limit cycles, rather than cliques.  Figure~\ref{fig:modular}B depicts a network that has 6 overlapping limit cycles, corresponding to subsets of nodes 1-5, 4-8, 7-11, 10-14, 13-17, and 16-20.  The network activity will stay in a single limit cycle indefinitely, unless it receives an external ``kick'' helping it to transition to an adjacent limit cycle.

\begin{figure}[!h]
\begin{center}
\includegraphics[width=6.15in]{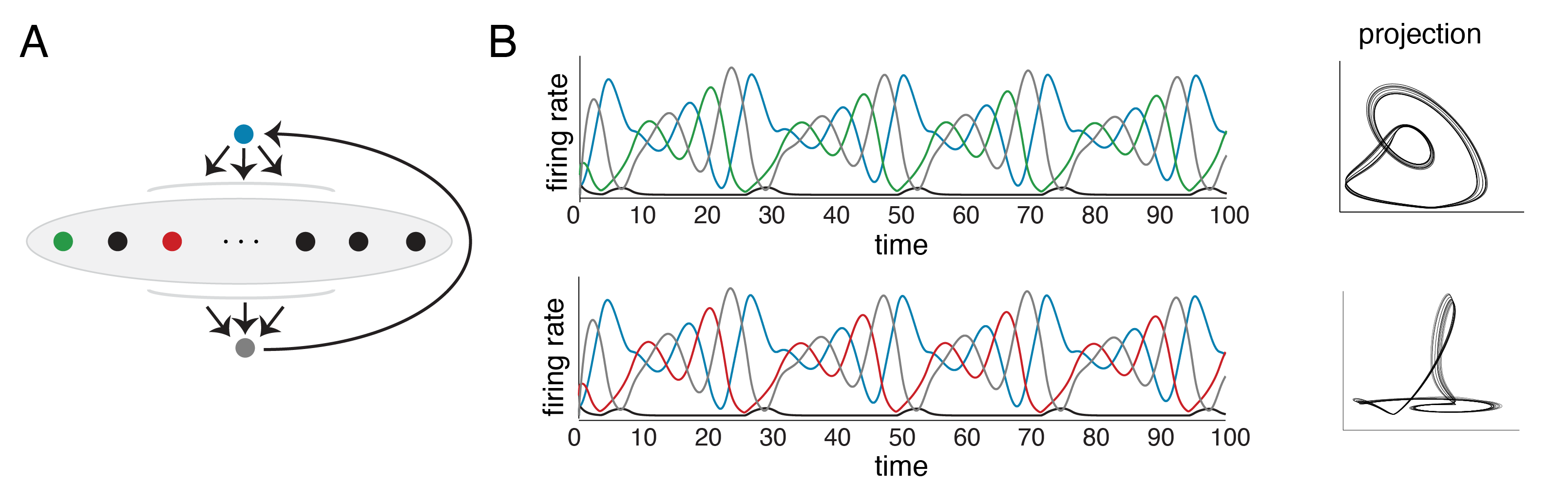}
\vspace{-.1in}
\end{center}
\caption{{\bf Multiple quasi-periodic attractors in the same network.} {(A)} A graph on $n$ nodes, where there are $n-2$ nodes in the middle (inner) layer.  The top (blue) outer node feeds onto all nodes in the middle layer, while the bottom (gray) outer node receives connections from all nodes in the middle layer and feeds back onto the top node.  This architecture produces $n-2$ different quasi-periodic attractors that each involve the top node, one of the middle nodes, and the bottom node.  {(B)}  Two distinct quasi-periodic attractors, one involving the green middle node (top), and the other involving the red middle node (bottom).  In each solution, the activity of all other middle nodes is small and synchronous, and is depicted in black.  Random projections of the activity (right) indicate that they are quasi-periodic trajectories and not perfect limit cycles.  %To create the top plot, the green node was initialized at $0.1$ and all others at $0$; for the bottom plot, the red node was initialized at $0.1$ and all others at $0$.  All simulations were run with $\theta=1$, $\varepsilon=0.25$, and $\delta=0.5$. 
}
\label{fig:quasiperiodic}
\end{figure}

Our next example shows that a single network, with very simple architecture, can have multiple quasi-periodic attractors.  The network in Figure~\ref{fig:quasiperiodic}A has $n$ nodes and $n-2$ quasi-periodic attractors.  Each quasi-periodic orbit selects a single node from the inner (shaded) region, which forms a dynamic sequence with the blue and gray outer nodes (Figure ~\ref{fig:quasiperiodic}B). 

\begin{figure}[!h]
\begin{center}
\includegraphics[width=6.15in]{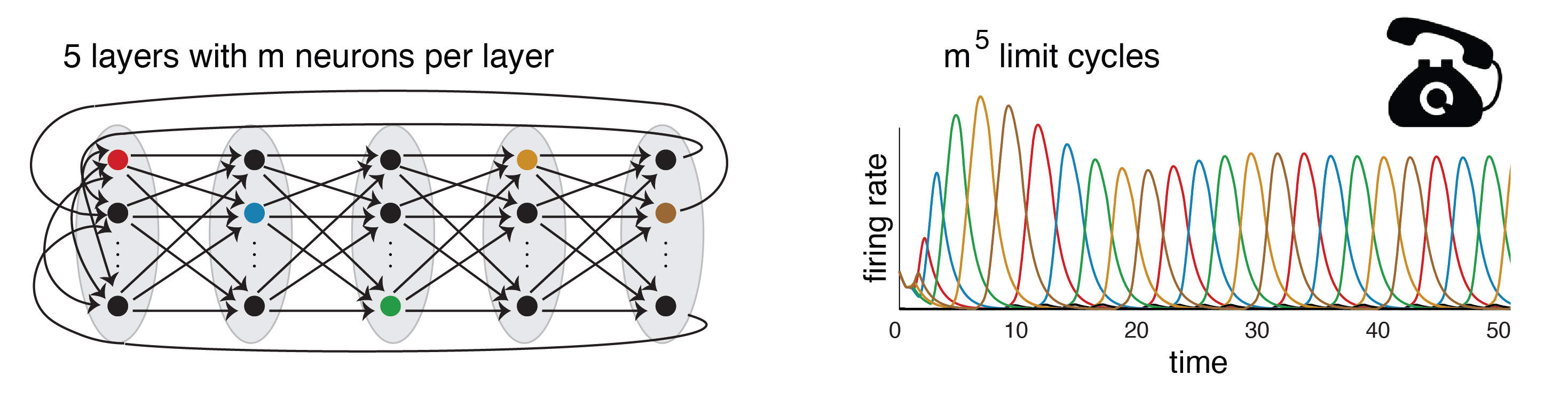}
\vspace{.1in}
\caption{{\bf The phone number network.} (Left) A cyclically structured graph with $m$ neurons per layer, and all $m^2$ feedforward connections from one layer to the next. (Right) A limit cycle for the corresponding CTLN (with parameters $\varepsilon=0.75$, $\delta=4$) in which 5 neurons fire in a repeating sequence consisting of one neuron from each layer.  Note there are $m^5$ such limit cycles, by symmetry.
}
\label{fig:phone-number}
\end{center}
\vspace{-.35in}
\end{figure}

Our last example, given in Figure~\ref{fig:phone-number}, is a generalization of the previous network -- now with $m$ nodes in each layer, all-to-all forward connections from one layer to the next, and the last layer connecting back to the first in a cyclic fashion. For parameters $\varepsilon=0.75$, $\delta=4$, the CTLN yields a limit cycle with sequential firing consisting of exactly one node from each layer, shown on the right.  By symmetry, there must be an equivalent limit cycle for every choice of 5 nodes, one from each layer, and thus the network is guaranteed to have $m^5$ limit cycles.  Note that this network architecture, increased to 7 layers, could serve as a mechanism for storing phone numbers in working memory (with $m = 10$ for digits $0-9$). The phone number is stored as a sequence that is repeated indefinitely, with different initial conditions producing different phone number sequences. 

%\FloatBarrier

\rkatie{\subsection{Discussion of CTLN dynamic attractors}}
\rkatie{We have seen a rich variety of emergent dynamics from CTLNs. Theorem~\ref{thm:Thm1b} gave us a simple way to guarantee that a network has dynamic attractors (as opposed to stable fixed points), by ensuring the graph is oriented and has no sinks. Alternatively, we can also engineer dynamic attractors from graphs with bidirectional edges, so that groups of neurons can fire synchronously as part of an evolving sequence of activity. For example, Theorem~\ref{thm:Thm2} tells us that if a clique has a target, then there is no associated fixed point within the larger network. However, such cliques can be transiently activated. This activity is typically followed by activation of the target node(s), which can in turn activate a subsequent clique, reminiscent of cell assembly sequences (see Figure~\ref{fig:modular}A). This kind of chain-like construction was also used in designing subnetworks for the quadruped gaits in Figure~\ref{fig:quadruped}.
}

\rkatie{In fact, the bound and trot graph architectures from Figure~\ref{fig:quadruped}A (left panel) are examples of a more general construction called a {\em cyclic union}. Cyclic unions are examples of ``gluing'' constructions, which combine subgraphs into a larger network according to certain rules. Specifically, in a cyclic union a set of disjoint graphs $G_{\tau_1}, \ldots, G_{\tau_N}$, is glued together in a cyclic manner such that every node in $G_{\tau_i}$ projects to every node in $G_{\tau_{i+1}}$, with no other edges between components. 
In Figure~\ref{fig:quadruped}A, the cyclic union has subgraphs corresponding to a $2$-clique, a singleton, another $2$-clique, and another singleton. Interestingly, the graphs in Figures~\ref{fig:quasiperiodic}A and~\ref{fig:phone-number} are also cyclic unions. In the case of the phone number network, the cyclic union consists of 5 independent sets of size $m$. Motivated in part by these examples, cyclic unions have been studied extensively in subsequent work \cite{fp-paper, sequential-att-paper}. For a summary of theorems about cyclic unions and other related ``gluing" constructions, see \cite{Notices-extended}.
}

\rkatie{When combining the two subnetworks for bound and trot gaits in Figure~\ref{fig:quadruped}A (right panel), we used a different method of ``gluing" two graphs. In this case, the two networks were combined by taking a union of all vertices and edges, while identifying the 4 common nodes (corresponding to LF, RF, LH, RH) across the two graphs. The mathematical properties of this type of gluing are not well understood. However, simulations have shown that this method often preserves the attractors of each component graph, allowing the construction of networks with multiple coexisting limit cycles. For example, in \cite{juliana-CPG} a similar process is used to construct a network with 5 different quadruped gaits, each accessible via different choices of initial conditions.
}

\vspace{.2in}

\section{Proofs}\label{sec:proofs}
In this section, we prove Theorems~\ref{thm:parity},~\ref{thm:Thm1} and~\ref{thm:Thm2} -- and, as a consequence, Theorem~\ref{thm:Thm1b} (an immediate corollary of Theorem~\ref{thm:Thm1}).  Our proofs build on our previous work in \cite{flex-memory, net-encoding, pattern-completion}, as well as a useful reframing of TLNs
as a patchwork of linear systems of ODEs, which we describe in Section~\ref{sec:linear-systems}.    

Throughout this section, TLNs will be assumed to be \rkatie{{\bf competitive} and } {\bf nondegenerate}, as originally defined in \cite{fp-paper}:

\begin{definition} \label{def:nondegenerate}
We say that a TLN $(W,b)$ is {\it nondegenerate} if 
\begin{itemize}
\item $\det(I-W_\sigma) \neq 0$ for each $\sigma \subseteq [n]$,
\item for each $\sigma \subseteq [n]$ such that $b_i >0$ for all $i \in \sigma$,  the corresponding Cramer's determinant is nonzero: $\det((I-W_\sigma)_i;b_\sigma) \neq 0$, and
\item $b_i > 0$ for at least one $i \in [n]$.
\end{itemize}
\end{definition}
As in previous sections, we use the notation $A_\sigma$ and $b_\sigma$ to denote a matrix $A$ and a vector $b$ that have been truncated to include only entries with indices in $\sigma$. 
Here we also use the notation $(A_i;b)$ to denote a matrix $A$ whose $i$th column has been replaced by the vector $b$, as in Cramer's rule (see e.g., Lemma 1 in \cite{fp-paper}).\footnote{The use of the subscript $i$ inside $(A_i;b)$ and $((A_\sigma)_i;b_\sigma)$ has a different meaning than the subscript $\sigma$ in $A_\sigma$, because it refers to replacing the $i$th column by $b$ (or $b_\sigma$), as opposed to restricting the entries of $A$ to the index set $\{i\}$. This is an abuse of notation, but the meaning should always be clear from the context.} In the case of a restricted matrix, $((A_\sigma)_i;b_\sigma)$ denotes the matrix $A_\sigma$ where the column corresponding to the index $i \in \sigma$ has been replaced by $b_\sigma$ (note that this is not typically the $i$th column of $A_\sigma$).
Note that almost all networks of the form~\eqref{eq:network} are nondegenerate, since having a zero determinant is a highly fine-tuned condition.   

\rkatie{The first nondegeneracy condition, $\det(I-W_\sigma) \neq 0$, ensures that all the linear systems of the TLN are nondegenerate, and thus have a unique fixed point.  The second condition is necessary to ensure that the fixed points of two adjacent linear systems do not coincide on a common boundary of their respective chambers in the state space.  Finally, the third condition guarantees that the origin is not a fixed point of the TLN. }

%%%%%%%%%%%%%%%%%%%%%%%%%%%%%%%%%

\subsection{TLNs as a patchwork of linear systems} \label{sec:linear-systems} 
Recall that for a general TLN $(W,b)$, the equations are given by:
$$\dfrac{dx_i}{dt} = -x_i + \left[\sum_{j=1}^n W_{ij}x_j+b_i \right]_+, \text{ for } i \in [n].$$
It is convenient to introduce the notation,
$$y_i(x) \od \sum_{j=1}^n W_{ij}x_j+b_i,$$
and rewrite the TLN equations as
$$\frac{dx_i}{dt}=-x_i + [y_i]_+, \text{ for } i \in [n].$$
From here we see that every TLN $(W,b)$ gives rise to a hyperplane arrangement with $n$ hyperplanes defined by $y_i(x) = 0$, subdividing $\RR^n$ into at most $2^n$ regions of the form:
$$R_\sigma=\{x\in \mathbb{R}^n~|~y_i(x) \geq 0 ~\forall~i \in \sigma, \text{ and } y_k(x) \leq 0 ~\forall~k \notin \sigma \}$$
Note that the regions cover $\RR^n$ and overlap only on boundaries where one or more $y_i=0$. Moreover, if $x^*$ is a fixed point of $(W,b)$, then
\begin{eqnarray}\label{eq:fp-eqs}
x_i^* = [y_i(x^*)]_+, \text{ for each } i \in [n].
\end{eqnarray}
Through this lens, we can view the piecewise-linear TLN dynamics as a patchwork of distinct linear systems, one for each $R_\sigma$. Restricting to $x \in R_\sigma$, we see that the ODE for each node $i \in \sigma$ reduces to ${dx_i}/{dt}=-x_i + y_i$, while for $k \not\in \sigma$ we obtain ${dx_k}/{dt}=-x_k$. The dynamics in each $R_\sigma$ thus reduce to a purely linear system, which we denote by $L_\sigma$:
\begin{center}
$L_\sigma= \left\{\dfrac{dx_i}{dt} = -x_i + \sum_{j=1}^n W_{ij}x_j+b_i \mid i \in \sigma\right\} ~\bigcup ~\left\{\dfrac{dx_k}{dt} = -x_k \mid k \notin \sigma\right\}. $
\end{center}
Assuming each linear system $L_\sigma$ is nondegenerate, which is guaranteed by the nondegeneracy condition for the TLN, each has a unique fixed point $x^*$ associated to it.  Note that $x^*$ is a fixed point of the TLN precisely when it lies inside the region $R_\sigma$ where its linear system $L_\sigma$ applies. 

\begin{figure}[!ht]
\begin{center}
\vspace{.2in}
\includegraphics[width=6.5in]{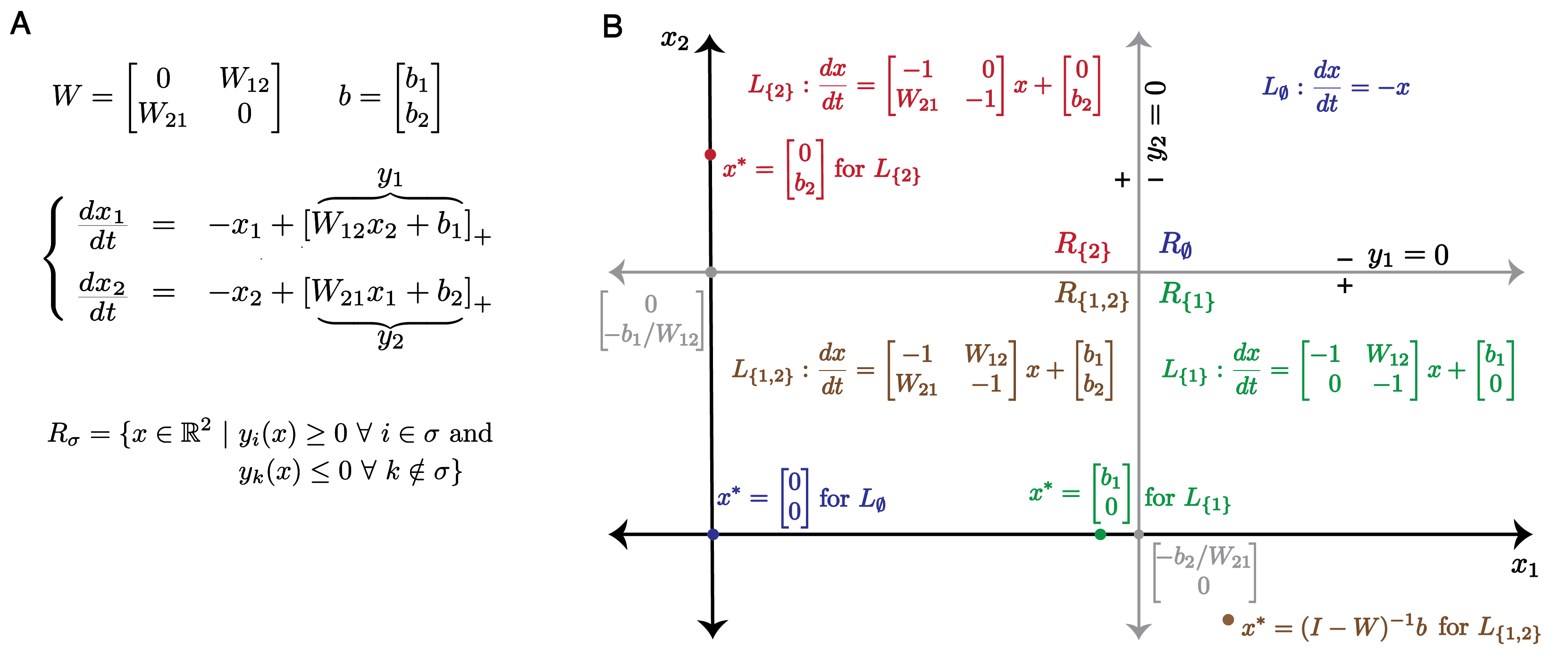}
\vspace{.1in}
\caption{{\bf Patchwork of linear systems for a TLN $\mathbf{(W,b)}$ on $\mathbf{n=2}$ nodes.} (A) Equations for a general $2 \times 2$ TLN and corresponding $R_\sigma$ regions. (B)  An arrangement of hyperplanes $y_i=0$ subdividing $\RR^2$ into regions $R_\sigma$, with corresponding linear systems $L_\sigma$. The fixed point of each linear system is also shown, color-coded according to its region.
}
\label{fig:linear-systems}
\end{center}
\vspace{-.2in}
\end{figure}

Figure~\ref{fig:linear-systems} illustrates the hyperplanes and regions for a TLN with $n=2$.  Each region, $R_\sigma$, has its own linear system of ODEs, $L_\sigma,$ for $\sigma = \emptyset, \{1\}, \{2\},$ or $\{1,2\}$. The fixed points corresponding to each linear system are denoted by $x^*$, in matching color. Note that only one of the four regions, $R_{\{2\}}$, contains its own fixed point (in red). This fixed point, $x^* = [0, b_2]^T$, is thus the only fixed point of the TLN.

As observed above, the regions $R_\sigma$ cover $\RR^n$, and within each $R_\sigma$ the TLN equations reduce to the linear system $L_\sigma$. In particular, this means that a fixed point $x^*$ of a TLN $(W,b)$ must be a fixed point of one of the linear systems, $L_\sigma$. Moreover, it must be an $L_\sigma$ fixed point that lies within the corresponding chamber, $R_\sigma$. To determine the fixed points of a TLN it is thus important to (1) identify the fixed points of the $L_\sigma$'s, and (2) have a method for checking whether they lie inside their respective chambers, $R_\sigma$. The following lemmas are useful for these purposes. 

\subsection{Fixed points of the linear systems $L_\sigma$}
\begin{lemma}\label{lemma:Lsigma}
Let $(W,b)$ be a competitive nondegenerate TLN on $n$ nodes. For each $\sigma \subseteq [n]$, the linear system $L_\sigma$ has a unique fixed point, $x^* \in \RR^n$, with $\supp x^* \subseteq \sigma$. This fixed point is {\it stable} if and only if $-I+W_\sigma$ is a stable matrix (that is, all eigenvalues have negative real part).
\end{lemma}

\begin{proof}
Fix $\sigma \subseteq [n]$. Without loss of generality, we can reorder the nodes so that $\sigma = \{1,\ldots,|\sigma|\}$. The linear system $L_\sigma$ is then given by
\begin{equation}\label{eq:Lsigma}
L_\sigma: \;\;\dfrac{dx}{dt} = Ax + \begin{bmatrix} b_\sigma \\ 0 \end{bmatrix}, \text{ for } \;
A = \begin{bmatrix} -I_{|\sigma|} + W_{\sigma}  & W_{\sigma \bar\sigma} \\ 
0 &  -I_{n-|\sigma|} \end{bmatrix},
\end{equation}
where $W_{\sigma \bar\sigma}$ is the matrix obtained by restricting $W$ to rows indexed in $\sigma$ and columns indexed by the complement, $\bar\sigma = [n]\setminus \sigma$. Note that $I_{|\sigma|}$ and $I_{n-|\sigma|}$ are identity matrices of sizes $|\sigma|$ and $n-|\sigma|$, respectively.

Now observe that $\det A = (-1)^{n-|\sigma|} \det(-I +W_{\sigma})$, where the size of $I$ is understood to match that of $W_\sigma$. By the nondegeneracy condition of Definition~\ref{def:nondegenerate}, we see that $\det A \neq 0$. It follows that $L_\sigma$ is nondegenerate and has a unique fixed point given by
$$x^* = (- A)^{-1} \begin{bmatrix} b_\sigma \\ 0 \end{bmatrix}.$$
(Note that $x^*$ may or may not lie inside the appropriate chamber $R_\sigma$.) Using Cramer's rule, we see that the entries of $x^*$ can be written as
\begin{eqnarray*}
x_i^* &=& \dfrac{\det\left((-A)_i;\begin{bmatrix} b_\sigma \\ 0 \end{bmatrix}\right)}{\det(-A)}
= \left\{ \begin{array}{cc} \dfrac{\det((I-W_\sigma)_i;b_\sigma)}{\det(I-W_\sigma)} & \text{for } i \in \sigma,\\
0 & \text{for } i \not\in \sigma. \end{array} \right.
\end{eqnarray*}
Thus, $\supp x^* \subseteq \sigma$. 
(Remark: In the case where $b_i>0$ for all $i \in \sigma$, the nondegeneracy condition guarantees that $x_i^* \neq 0$ for $i \in \sigma$. However, this does not guarantee that $x_i^* > 0$; it could be positive or negative.)

The fixed point $x^*$ is {\it stable} if and only if the Jacobian of the system evaluated at $x^*$ is stable. In a linear system of the form $dx/dt = Ax + b$, the Jacobian at every point is simply the matrix $A$. In our case, the matrix $A$ for the system $L_\sigma$ is block triangular, so the eigenvalues of $A$ are the eigenvalues of $-I+W_\sigma$ together with $n-|\sigma|$ copies of $-1$. It follows that the fixed point $x^*$ is stable if and only if the eigenvalues of $-I+W_\sigma$ all have negative real part. In other words, $x^*$ is stable if and only if $-I+W_\sigma$ is a stable matrix.
\end{proof}

As an immediate corollary, we see that our TLNs have at most one fixed point per $\sigma \subseteq [n]$:

\begin{corollary}\label{cor:fp-bound}
A competitive nondegenerate TLN on $n$ nodes has at most one fixed point per linear system $L_\sigma$, for $\sigma \subseteq [n]$. In particular, there are at most $2^n$ fixed points.
\end{corollary}

\subsection{From $L_\sigma$ fixed points to TLN fixed points}
Whether or not the fixed point $x^*$ of the linear system $L_\sigma$ is actually a fixed point of the full TLN depends on whether or not $x^* \in R_\sigma$, as this is the region where the $L_\sigma$ equations apply. By definition, $x^* \in R_\sigma$ if and only if 
\begin{itemize}
\item $y_i(x^*) \geq 0$ for all $i \in \sigma$, and
\item $y_k(x^*) \leq 0$ for all $k \not\in \sigma$. 
\end{itemize}
These are known as the ``on"-neuron and ``off"-neuron conditions, respectively. 
Recall that for $i \in \sigma$, the $L_\sigma$ equations are:
\begin{eqnarray*}
\dfrac{dx_i}{dt} &=& -x_i + y_i(x), \text{ for all } i \in \sigma,\\
\dfrac{dx_k}{dt} &=& -x_k, \text{ for all } k \not\in \sigma.
\end{eqnarray*}
It follows that at the $L_\sigma$ fixed point,
$$x_i^* = y_i(x^*) \text{ for all } i \in \sigma, \text{ and } x_k^* = 0 \text{ for all } k \not\in \sigma.$$
On the other hand, for $x^*$ to be a fixed point of the TLN it must satisfy $x_j^* = [y_j(x^*)]_+$ for all $j \in [n]$ (see equation~\eqref{eq:fp-eqs}).
The ``on'' conditions thus ensure that $x_i^* \geq 0$ for all $i \in \sigma$, while the ``off'' conditions
guarantee that $x_k^* = 0$ satisfies $x_k^* = [y_k(x^*)]_+$, for each $k \notin \sigma$.
We thus have the following equivalent formulation for when a fixed point $x^*$ of $L_\sigma$ is a fixed point of the full TLN  \cite{book-chapter}:
\begin{itemize}
\item (``on'' conditions) \: $x_i^* \geq 0$ for all $i \in \sigma$, and
\item (``off'' conditions) \: $y_k(x^*) \leq 0$ for all $k \not\in \sigma$. 
\end{itemize}

The next lemma gives precise formulas for checking the ``on''- and ``off''-neuron conditions. It also shows that if all $b_j > 0$, the inequalities are strict. This means that when the conditions are satisfied, $\supp x^* = \sigma$.

\begin{lemma}\label{lemma:y_i(x^*)}
Consider a competitive nondegenerate TLN $(W,b)$ on $n$ nodes, and let $x^*$ be the fixed point of $L_\sigma$ for some $\sigma \subseteq [n]$. Then
\begin{eqnarray*}
y_i(x^*) &=& x_i^* = \dfrac{\det((I-W_\sigma)_i;b_\sigma)}{\det(I-W_\sigma)}, \text{ for all } i \in \sigma, \text{ and}\\
y_k(x^*) &=& \dfrac{\det((I-W_{\sigma \cup \{k\}})_k;b_{\sigma \cup \{k\}})}{\det(I-W_\sigma)},
\text{ for all } k \not\in \sigma.
\end{eqnarray*}
Moreover, if $b_j > 0$ for all $j \in [n]$, then $y_i(x^*) = x_i^* \neq 0$ for all $i \in \sigma$ and $y_k(x^*) \neq 0$ for all
$k \notin \sigma$.
\end{lemma}

\begin{proof}
Let $x^*$ be the fixed point of $L_\sigma$. We have already seen that $y_i(x^*) = x_i^*$ for $i \in \sigma$. The expression for $x_i^*$ in the statement is simply the Cramer's rule computation for $x_i^*$ from the proof of Lemma~\ref{lemma:Lsigma}. Moreover, as explained in the remark at the end of that proof, if $b_j > 0$ for all $j \in [n]$, then the nondegeneracy condition implies $x_i^* \neq 0$ 
for all $i \in \sigma$. 

To see the $y_k(x^*)$ equation for $k \notin \sigma$, we use the definition of $y_k(x)$ and the formula for $x_i^*$ in the proof of Lemma~\ref{lemma:Lsigma} (for both $i \in \sigma$ and $i \notin \sigma$)  to compute:
\begin{eqnarray*}
y_k(x^*) &=& \sum_{j=1}^n W_{kj}x^*_j + b_k 
= \sum_{j \in \sigma} W_{kj} \dfrac{\det((I-W_\sigma)_j;b_\sigma)}{\det(I-W_\sigma)} + b_k.
\end{eqnarray*}
Now consider the Cramer's determinant,
$\det((I-W_{\sigma \cup \{k\}})_k;b_{\sigma \cup \{k\}})$, where the matrix $((I-W_{\sigma \cup \{k\}})_k;b_{\sigma \cup \{k\}})$ has the vector $b_{\sigma \cup \{k\}}$ in the column corresponding to index $k$ of $I-W_{\sigma \cup \{k\}}$. 
We can move the column corresponding to index $k$ to the beginning with some number of column swaps. An equal number of row swaps brings the row corresponding to index $k$ to the top. We thus obtain,
$$\det((I-W_{\sigma\cup \{k\}})_k;b_{\sigma\cup \{k\}})  =
 \det\left(\begin{array}{c|c} b_k & -W_{ki_1} \cdots -W_{ki_{|\sigma|}}\\ \hline b_{\sigma} & I-W_{\sigma}  \end{array}\right),$$
 where $\sigma = \{i_1,\ldots,i_{|\sigma|}\}$. Applying the Laplace expansion for the determinant along the first row, we compute
\begin{eqnarray*}
\det((I-W_{\sigma\cup \{k\}})_k;b_{\sigma\cup \{k\}})  &=& b_k \det(I-W_\sigma) + \sum_{\ell = 1}^{|\sigma|} (-1)^{2+\ell} (-W_{k i_\ell}) (-1)^{\ell-1} \det((I-W_\sigma)_\ell; b_\sigma)\\
 &=& b_k \det(I-W_\sigma) + \sum_{\ell = 1}^{|\sigma|} W_{k i_\ell} \det((I-W_\sigma)_\ell; b_\sigma)\\
  &=& \sum_{j \in \sigma} W_{k j} \det((I-W_\sigma)_j; b_\sigma) + b_k \det(I-W_\sigma).
\end{eqnarray*}
Therefore, $y_k(x^*) = \dfrac{\det((I-W_{\sigma \cup \{k\}})_k;b_{\sigma \cup \{k\}})}{\det(I-W_\sigma)},$ as desired.
\end{proof}

An immediate corollary of Lemma~\ref{lemma:y_i(x^*)} is that none of the fixed points of the linear systems $L_\sigma$ lie on any of the hyperplanes defined by $y_i = 0$. In particular, we obtain the following:

\begin{corollary} \label{cor:Rsigma}
Let $(W,b)$ be a competitive nondegenerate TLN with $b_j>0$ for all $j \in [n]$.
If $x^*$ is a fixed point of $(W,b)$, then $x^*$ lies in the interior of $R_\sigma$ for $\sigma = \supp x^*$. 
In particular, the fixed points of $(W,b)$ are all isolated.
\end{corollary}

It is worth noting that the guarantee of isolated fixed points relies crucially on the TLN being nondegenerate.  In the absence of nondegeneracy, it is possible to have two fixed points coincide with each other on a boundary $y_i=0$.

%%%%%%%%%%%%%%%%%%%%%%%%%%%%%%%%%
\subsection{Proof of Theorem~\ref{thm:parity}}\label{sec:parity-proof}
The proof of Theorem~\ref{thm:parity} is a straightforward application of the Poincar\'e-Hopf theorem, the famous index theorem relating the zeros of a vector field on a manifold $\M$ to the Euler characteristic $\chi(\M)$.  First we define the index of a zero of a vector field, following the notation of \cite{Milnor1965}.

\begin{definition}\label{def:index-vec-field} 
Let $\M$ be an oriented differentiable manifold.  Assume that a continuous vector field $v(x)$ on $\M$ is differentiable in an open neighborhood of a zero, $x^* \in v^{-1}(0)$. If the vector field's Jacobian ${\frac{dv}{dx}(x^*)}$ has all nonzero eigenvalues, then the index $ \iota(v, x^*)$ is defined as the sign of its determinant: 
\begin{equation}  
\iota(v, x^*)=\sgn \det  \left(\frac{dv}{dx}(x^*)\right).
\end{equation}
\end{definition}

\begin{theorem}[Poincar\'e-Hopf theorem \cite{Milnor1965}]\label{thm:Poincare-Hopf} 
Let $\M$ be a compact oriented differentiable manifold with boundary $\partial \M$, and let $v(x)$ be a continuous vector field on $\M$ with isolated zeros.  Assume that at each point of the boundary, the vector field is pointing outside of $\M$. Then the following equality holds:
\begin{equation}\label{eq:PoincareHopf} 
\sum_{x^*\in v^{-1}(0)} \iota(v,x^*)=\chi(\M). 
\end{equation}
\end{theorem} 

\begin{figure}[!h]
\begin{center}
\includegraphics[width=2.75in]{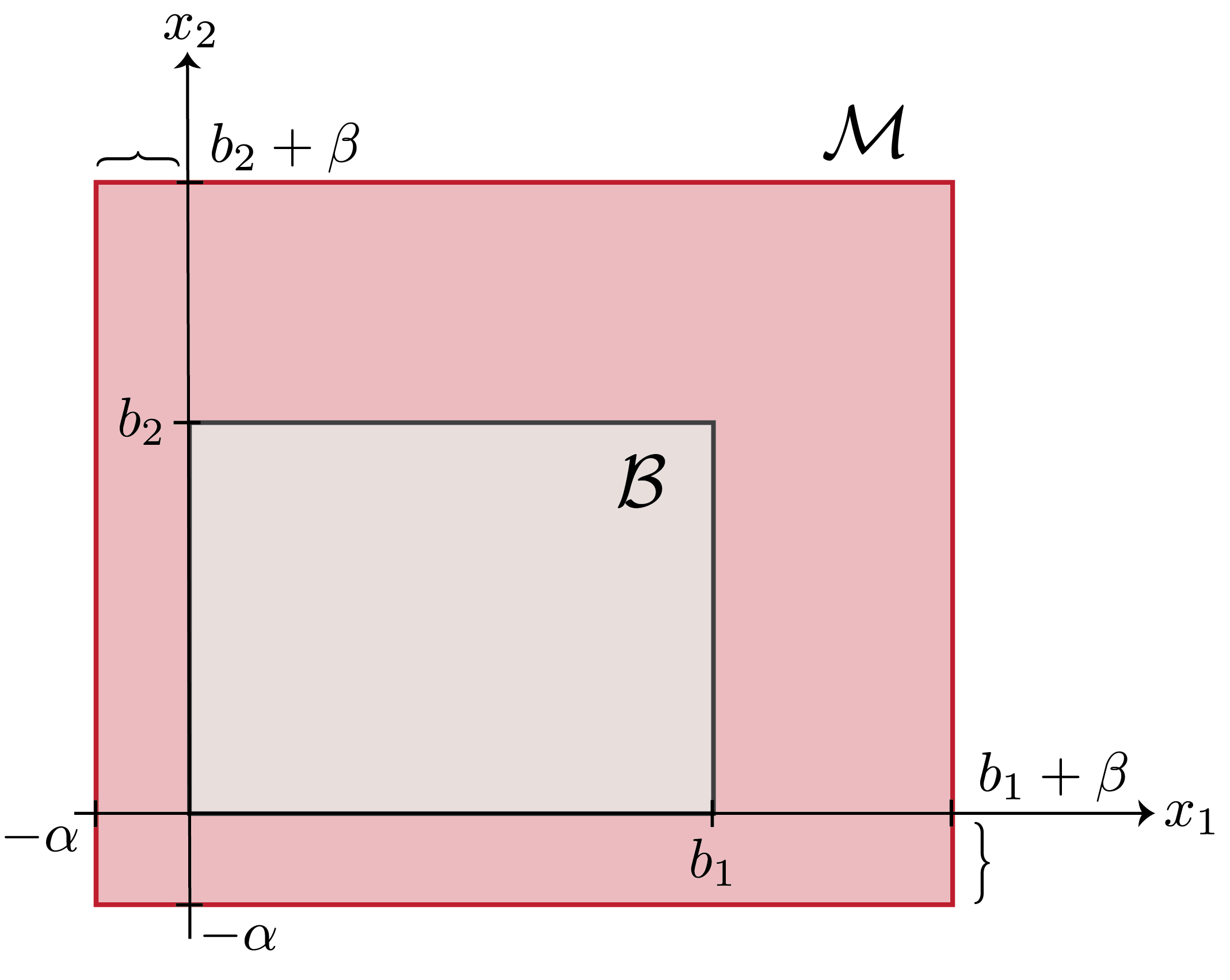}
\vspace{.1in}
\caption{{\bf A two-dimensional schematic of the hypercube $\mathbf{\M \subset \RR^n}$}. By design, $\M$ contains the box $\B =  \prod_{i=1}^n [0,b_i]$ from Lemma~\ref{lemma:bounded}. $\M$ is used in the proof of Theorem~\ref{thm:parity}.}
\label{fig:M-box}
\end{center}
\vspace{-.1in}
\end{figure}

To apply the Poincar\'e-Hopf theorem in the context of competitive nondegenerate TLNs, we will consider the vector field $v(x)=x-[Wx+b]_+$, which is the negative of the vector field in equation~\eqref{eq:network}. The zeros of $v(x)$ are thus precisely the fixed points of the TLN. The manifold of interest will be a hypercube $\M \subset \RR^n$ that contains the box $\B = \prod_{i=1}^n [0,b_i]$ from Lemma~\ref{lemma:bounded} (see Figure~\ref{fig:M-box}). It will thus contain all the fixed points of the TLN (Corollary~\ref{cor:box-fp}). The reason we choose $v(x)$ to have the opposite sign of the TLN vector field is so that it points {\it outside} of $\M$ on the boundary, matching the hypothesis in Theorem~\ref{thm:Poincare-Hopf}. In contrast, the TLN vector field, $-v(x)$, points in, yielding bounded activity with $\B\subset \M$ a globally attracting set (Lemma~\ref{lemma:bounded}).

For convenience, we restate Theorem~\ref{thm:parity} below, and then proceed to the proof. Recall that the index we defined for fixed points of TLNs is $\idx(\sigma) = \sgn \det (I-W_\sigma)$, where $\sigma$ is the fixed point support. It is guaranteed to be nonzero for nondegenerate TLNs.\\

\noindent {\bf Theorem~\ref{thm:parity}} (parity).  
Let $(W,b)$ be a competitive nondegenerate TLN on $n$ nodes, with $b_i>0$ for all $i \in [n]$. Then
$$\sum_{\sigma \in \FP(W,b)} \idx(\sigma) = + 1.$$
In particular, the total number of fixed points $|\FP(W,b)|$ is always odd.

\begin{proof}
Let $x^*$ be a fixed point of a competitive nondegenerate TLN $(W,b)$, with $b_i > 0$ for all $i \in [n]$, and let $v(x)=x-[Wx+b]_+$ be the negative of the TLN vector field.
First, we will show that the index $\iota(v,x^*)$ from the Poincar\'e-Hopf theorem is equal to the TLN index, 
$\idx(\sigma),$ for $\sigma = \supp x^*$. Starting from Definition~\ref{def:index-vec-field},
$$\iota(v, x^*)=\sgn \det  \left(\frac{dv}{dx}(x^*)\right) = \sgn \det(-A) = \sgn \det(I-W_\sigma) = \idx(\sigma),$$
where $A$ is the matrix for $L_\sigma$ in equation~\eqref{eq:Lsigma}. Because $x^*$ lies in the interior of $R_\sigma$ (Corollary~\ref{cor:Rsigma}), where the dynamics are governed by the linear system $L_\sigma$, the matrix $A$ is precisely the Jacobian of the TLN vector field at $x^*$. It follows that $-A$ is the Jacobian of $v$ evaluated at $x^*$.

Let $\M$ be the hypercube 
$$\M=\prod_{i=1}^n [-\alpha,~ b_i + \beta],$$
for some $\alpha,\beta>0$, and observe that $\M$ contains $\B = \prod_{i=1}^n [0, b_i]$ (see Figure~\ref{fig:M-box}). Thus, by Corollary~\ref{cor:box-fp}, $\M$ contains all fixed points of $(W,b)$ in its interior. Additionally, $\M$ is compact, oriented, and has a boundary $\partial \M$. Note that although $\M$ is not differentiable on some parts of the boundary (it has corners), the vector field $v(x)$ is continuous on all of $\M$ and has isolated zeros in the interior (Corollary~\ref{cor:Rsigma}).\rkatie{\footnote{\rkatie{The boundary of $\M$ could easily have been chosen to be smooth, so that $\M$ is differentiable everywhere. However, we pick a hybercube for ease of description and showing that $v(x)$ satisfies the additional property of pointing outside of $\M$ along the boundary, as required by Theorem~\ref{thm:Poincare-Hopf}. (By continuity, $v(x)$ also points outside the boundary of a perturbed, smoothed version of $\M$.)}}}

To show that at each point on the boundary, $\partial M$, $v(x)$ points outside of $\M$, we will now make some convenient choices for $\alpha, \beta$. Let $w \od \max_{i,j} |W_{ij}|$. Fix some $\alpha > 0$, and let $\beta = n\alpha w.$
Note that $\M$ is bounded by $2n$ hyperplanes: $n$ hyperplanes of the form $x_i=-\alpha$, with outward pointing normal vectors $-e_i$; and $n$ hyperplanes of the form $x_i=b_i + n\alpha w$, with outward pointing normal vectors $e_i$.  To show that $v(x)$ points outward everywhere along $\partial \M$, it thus suffices to show that on each of the $x_i=-\alpha$ hyperplanes we have $v_i(x)<0$, and on each of the $x_i=b_i+ n\alpha w$ hyperplanes we have $v_i(x)>0$.  

Recall that the $i$th coordinate of $v(x)$ is given by $v_i(x) = x_i -  [\sum_{j=1}^n W_{ij}x_j + b_i]_+$.
When $x_i= -\alpha$, we have $v_i(x) = -\alpha - [\sum_{j=1}^n W_{ij}x_j + b_i]_+ < 0$, as desired.  On the opposite hyperplane, where $x_i=b_i+ n\alpha w$, we have $v_i(x) = b_i + n\alpha w -  [\sum_{j=1}^n W_{ij}x_j + b_i]_+ \geq~ b_i+ n\alpha w - \max_{x \in \partial M} [\sum_{j=1}^n W_{ij}x_j + b_i]_+$.  Since $W_{ij} \leq 0$ and $W_{ii}=0$, the largest value of $[\sum_{j=1}^n W_{ij}x_j + b_i]_+$ on $\partial M$ occurs when $x_j = -\alpha$ for each $j \neq i$. We thus have
\begin{eqnarray*}
v_i(x) &\geq& b_i + n\alpha w - \left[\sum_{j=1}^n W_{ij}(-\alpha) + b_i\right]_+ 
\geq b_i + n\alpha w - \left[\sum_{j\neq i} (-w)(-\alpha) + b_i\right]_+ \\
&=& b_i + n\alpha w - ((n-1) \alpha w + b_i) = \alpha w > 0,
\end{eqnarray*}
where the second inequality stems from the fact that the most negative value of $W_{ij}$ is $-w$.
We conclude that the vector field $v(x)$ points outward everywhere along the boundary of $\M$, and thus the Poincar\'e-Hopf theorem applies.

Now recall that the Euler characteristic $\chi(\M) = 1$ for any solid hypercube (or ball) in $\RR^n$. It follows from Theorem~\ref{thm:Poincare-Hopf} that
$$\sum_{\sigma \in \FP(W,b)} \idx(\sigma) = \sum_{x^*\in v^{-1}(0)} \iota(v,x^*) = \chi(\M) = + 1.$$
\end{proof}

\subsection{Proof of Theorem~\ref{thm:Thm1}}\label{sec:proof-Thm1}

In this section, we prove Theorem~\ref{thm:Thm1}.  
Recall from Lemma~\ref{lemma:Lsigma} that a fixed point $x^*$ with support $\sigma$ is {\it stable} if and only if $-I+W_\sigma$ is a stable matrix (i.e., all eigenvalues have strictly negative real part). We say that a matrix is {\it unstable} if at least one eigenvalue has positive real part. The following lemma gives a condition on $W$ that rules out the existence of stable fixed points supported on two or more nodes.

\begin{lemma} \label{lemma:2x2}
Consider a TLN $(W,b)$ where $W$ has diagonal entries $W_{ii} = 0$.  If all $2 \times 2$ principal submatrices of $-I+W$ are unstable, then the network has no stable fixed points supported on more than one node.
\end{lemma}

\begin{proof}
Suppose all $2 \times 2$ principal submatrices, $-I+W_\tau$ for $|\tau| = 2$, are unstable, and note that the trace of each of these matrices is $-2$. In \cite[Lemma 1]{net-encoding}, it was shown that if all $2 \times 2$ principal submatrices of an $n \times n$ matrix have negative trace and are unstable, then all larger principal submatrices are also unstable.  It follows that $-I+W_\sigma$ is unstable for all $\sigma$ of size $|\sigma| \geq 2$. 
This implies that $(W,b)$ can have no stable fixed points supported on more than one node, because a fixed point  with support $\sigma$ can only be stable if $-I+W_\sigma$ is stable (Lemma~\ref{lemma:Lsigma}).
\end{proof}

\noindent We are now ready to prove Theorem~\ref{thm:Thm1}, which we restate below for convenience.\\

\noindent {\bf Theorem~\ref{thm:Thm1}.} Consider a competitive \rkatie{nondegenerate} threshold-linear network with connectivity matrix $W$, associated graph $G_W$, and uniform inputs $b_i=\theta$.  Suppose that:
\begin{itemize}
\item[(i)] $G_W$ is an oriented graph with no sinks, and
\item[(ii)] whenever $j \rightarrow i$ in $G_W$, $W_{ij} < \dfrac{1}{W_{ji}}$.
\end{itemize}
Then the network~\eqref{eq:network} has no stable fixed points.  Moreover, the network activity is bounded.

\begin{proof}
We have already seen that the activity is bounded (Lemma~\ref{lemma:bounded}).  To see that there are no stable fixed points, we first show that there can be no stable fixed points supported on two or more nodes.  By Lemma~\ref{lemma:2x2}, it suffices to show that all $2\times 2$ principal submatrices of $-I + W$ are unstable.  Each of these matrices, $\left(\begin{array}{cc} -1 & W_{ij} \\ W_{ji} & -1 \end{array}\right),$ has negative trace, and is thus stable if and only if its determinant is positive.  The determinant is
$$\Delta = \det\left(\begin{array}{cc} -1 & W_{ij} \\ W_{ji} & -1 \end{array}\right) = 1-W_{ij}W_{ji},$$
which is positive if and only if $W_{ij}W_{ji} < 1$.  

Since the network is competitive and $G_W$ is an oriented graph, there are two cases to consider: (a) there is a single edge $j \to i$ (or $i \to j$), so that $-1<W_{ij} \leq 0$ and $W_{ji} \leq -1$ (or vice versa), or (b) there is no edge between $i$ and $j$, so that both $W_{ij} \leq -1$ and $W_{ji} \leq -1$.  In case (a), hypothesis (ii) of the theorem implies $W_{ij}W_{ji} > 1$ (note that the inequality reverses, since $W_{ji}$ is negative).  We thus have $\Delta < 0$, and the $2 \times 2$ matrix is unstable.  In case (b), we immediately see that $W_{ij}W_{ji} \geq 1$ and thus $\Delta \leq 0$.  We conclude that all $2\times 2$ principal submatrices of $-I + W$ are unstable, and thus the network has no stable fixed points with 2 or more active nodes.

Next, we show by contradiction that the network has no fixed points supported on a single node (i.e., there is no winner-take-all behavior).  Suppose $x^*$ is a fixed point supported on node $i$, so that $x_i^*>0$ and $x_j^* = 0$ for all $j \neq i$, and recall that  $W_{ii} = 0$.  It follows that
$$x_i^* = \left[\sum_{j=1}^n W_{ij}x_j^*+\theta\right]_+ = [W_{ii} x_i^* + \theta]_+ = \theta.$$
On the other hand, for any $k \neq i$ we must have $x_k^* = 0$ at the fixed point, and so
$$x_k^* = \left[\sum_{j=1}^n W_{kj}x_j^*+\theta\right]_+ =[W_{ki}x_i^*+\theta]_+= [W_{ki}\theta+\theta]_+ = 0.$$
Now recall that $G_W$ has no sinks (by hypothesis (i) of the theorem), and so there exists at least one vertex $\ell \neq i$ such that $i \rightarrow \ell$.  This means $W_{\ell i} > -1$, and thus $x_\ell^* > 0$, contradicting the assumption that the fixed point was supported only on node $i$.
\end{proof}

\subsection{Proof of Theorem~\ref{thm:Thm2}}

To prove Theorem~\ref{thm:Thm2}, we make use of the fixed point conditions that were derived in~\cite{pattern-completion}.

\begin{lemma}\label{lemma:CTLN}
Consider \rkatie{a nondegenerate} CTLN model with $W = W(G,\varepsilon, \delta)$, and suppose $x^*$ is a fixed point of~\eqref{eq:network} supported on a clique $\sigma$ of $G$.  Then $x^*$ is a stable fixed point, and
$$x_\sigma^* = \dfrac{\theta}{\varepsilon+(1-\varepsilon)|\sigma|}1_\sigma,$$
where $1_\sigma$ is a column vector of all $1$s.
\end{lemma}

\begin{proof}
If $\sigma$ is a clique of $G$, then $-I+W_\sigma = (-1+\varepsilon)11^T - \varepsilon I_\sigma,$
with eigenvalues $|\sigma|(-1+\varepsilon)-\varepsilon$ and $-\varepsilon$.  Clearly, these are negative
for $0<\varepsilon<1$, so we can conclude that $-I+W_\sigma$ is stable.  It follows that any fixed point $x^*$ with support $\sigma$ is stable and unique \cite[Corollary 9]{net-encoding} (see also \cite[Section 1.1]{pattern-completion}).  To verify the formula for $x_\sigma^*$, we simply check that it satisfies the fixed point equation $x_\sigma^* = [W_\sigma x_\sigma^* + \theta 1_\sigma]_+.$
 Since $x_\sigma^*>0$, we can drop the threshold nonlinearity to obtain the equivalent constraint,
$(I-W_\sigma)x_\sigma^* = \theta 1_\sigma.$  Now plugging in the desired expression for $x_\sigma^*$ yields:
$$(I-W_\sigma)x_\sigma^* = \left((1-\varepsilon)11^T + \varepsilon I_\sigma\right) \dfrac{\theta}{\varepsilon+(1-\varepsilon)|\sigma|}1_\sigma = \dfrac{\theta}{\varepsilon+(1-\varepsilon)|\sigma|} ((1-\varepsilon)|\sigma|1_\sigma + \varepsilon 1_\sigma) = \theta 1_\sigma.$$
\end{proof}

\noindent We can now prove Theorem~\ref{thm:Thm2}, which we restate below for convenience.\\

\noindent {\bf Theorem~\ref{thm:Thm2}}. Let $G$ be a simple directed graph, and consider an associated  \rkatie{nondegenerate} CTLN with $W = W(G,\varepsilon, \delta)$ for any choice of the parameters $\varepsilon, \delta, \theta >0$ with $\varepsilon < 1.$
If $\sigma$ is a clique of $G$, then there exists a stable fixed point with support $\sigma$ if and only if $\sigma$ is target-free.

\begin{proof}
($\Rightarrow$) Suppose $x^*$ is a stable fixed point with support $\sigma$, where $\sigma$ is a clique of $G$.
Then  $W_{ij} = -1+\varepsilon$ for all pairs $i,j \in \sigma$.  To see that $\sigma$ must be a {\it target-free} clique, suppose that $\sigma$ has a target $k \notin \sigma$.  This implies that $W_{ki} = -1+\varepsilon$ for each $i \in \sigma$.  It follows that
$$x_k^* = \left[\sum_{i \in \sigma} W_{ki} x_i^*+ \theta\right]_+ =  
\left[(-1+\varepsilon) \sum_{i \in \sigma} x_i^*+ \theta\right]_+ = 
\left[\dfrac{\varepsilon \theta}{\varepsilon + (1-\varepsilon)|\sigma|}\right]_+>0,$$
where we have used the expression for $x_\sigma^*$ from Lemma~\ref{lemma:CTLN} to obtain
$\sum_{i \in \sigma} x_i^* = \dfrac{\theta |\sigma|}{\varepsilon+(1-\varepsilon)|\sigma|}.$
This contradicts the fact that $x_k^* = 0$, since $ k \notin \sigma$ and $\sigma = \supp x^*$.  We thus conclude that
$\sigma$ must be a target-free clique.

($\Leftarrow$) Suppose $\sigma$ is a target-free clique.  Since $\sigma$ is a clique, it follows from Lemma~\ref{lemma:CTLN} that if a fixed point $x^*$ with support $\sigma$ exists, then it must be unique and stable, with $x_\sigma^* = \dfrac{\theta}{\varepsilon+(1-\varepsilon)|\sigma|}1_\sigma.$ 
Clearly, $x_\sigma^*>0$, and thus the ``on''-neuron conditions hold. To guarantee that the fixed point of the CTLN with support $\sigma$ exists, however, we must also check
the ``off''-neuron conditions: $y_k(x^*) \leq 0 $ for each $k \notin \sigma$.
Since $\sigma$ is a target-free clique, for any $k \notin \sigma$ there exists $i_k \in \sigma$ such that $i_k \not\to k$, and so $W_{k i_k} = -1-\delta$.  We thus have
$$y_k(x^*) = \sum_{i \in \sigma} W_{ki} x_i^*+ \theta \leq \theta\left(\dfrac{-1 - \delta + (|\sigma|-1)(-1+\varepsilon)}{\varepsilon + (1-\varepsilon)|\sigma|}+1\right) = \dfrac{-\theta \delta}{\varepsilon + (1-\varepsilon)|\sigma|}<0,$$
showing that the ``off'' conditions hold.
\end{proof}

\vspace{.25in}

\section{Acknowledgments}
KM was supported by NIH R01 EB022862 and NSF DMS-1951599.  
VI was supported by NSFIOS-155925 and the NSF Next Generation Networks for Neuroscience Program (award 2014217).
CC was supported by NIH R01 EB022862, NSF DMS-1951165, and NSF DMS-1516881.  
CC, KM, and VI also gratefully acknowledge the support of the Statistical and Applied Mathematical Sciences Institute, under grant NSF DMS-1127914.

%--------------
% Bibliography
%--------------

\bibliographystyle{unsrt}
\bibliography{CTLN-refs}

\end{document}